\documentclass[aps,prx,reprint,preprintnumbers,superscriptaddress,nofootinbib,longbibliography,floatfix]{revtex4-2}
\pdfoutput=1
\usepackage{rotating}
\usepackage{array}
\usepackage{amsmath}
\usepackage[normalem]{ulem}
\usepackage{slashed}
\usepackage{booktabs}
\usepackage[pdftex,table]{xcolor}
\usepackage{units}
\usepackage{xfrac}
\usepackage{mathtools}
\usepackage{empheq}
\usepackage[]{units}
\usepackage{multirow}
\usepackage{amssymb}
\usepackage{url}
\usepackage{comment}
\usepackage{physics}
\usepackage{color,soul}
\usepackage{bbm}
\usepackage[caption=false]{subfig}
\usepackage{adjustbox}

\usepackage{hyperref}
\hypersetup{
  colorlinks=true,
  citecolor=blue,
  linkcolor=blue,
  urlcolor=blue
}

\newcommand{\pt}{\mathrm{p_{T}}}

\begin{document}

\title{Fast Point Cloud Generation with Diffusion Models in High Energy Physics}

\author{Vinicius Mikuni}
\email{vmikuni@lbl.gov}
\affiliation{National Energy Research Scientific Computing Center, Berkeley Lab, Berkeley, CA 94720, USA}

\author{Benjamin Nachman}
\email{bpnachman@lbl.gov}
\affiliation{Physics Division, Lawrence Berkeley National Laboratory, Berkeley, CA 94720, USA}
\affiliation{Berkeley Institute for Data Science, University of California, Berkeley, CA 94720, USA}

\author{Mariel Pettee}
\email{mpettee@lbl.gov}
\affiliation{Physics Division, Lawrence Berkeley National Laboratory, Berkeley, CA 94720, USA}

\begin{abstract}
    Many particle physics datasets like those generated at colliders are described by continuous coordinates (in contrast to grid points like in an image), respect a number of symmetries (like permutation invariance), and have a stochastic dimensionality.  For this reason, standard deep generative models that produce images or at least a fixed set of features are limiting.  We introduce a new neural network simulation based on a diffusion model that addresses these limitations named Fast Point Cloud Diffusion (FPCD).  We show that our approach can reproduce the complex properties of hadronic jets from proton-proton collisions with competitive precision to other recently proposed models.  Additionally, we use a procedure called progressive distillation to accelerate the generation time of our method, which is typically a significant challenge for diffusion models despite their state-of-the-art precision.
\end{abstract}

\maketitle


\section{Introduction}
\label{sec:intro}

Simulations are a critical component of nearly all inference tasks in particle physics.  These simulations connect theory to experiment and must span a wide range of energy scales and encode the complex structure of high energy physics data.  Physics-based simulations are excellent, but are only an approximation to nature.  Additionally, some components of these simulations are computationally expensive and are a bottleneck for the high statistics datasets that are being collected now and in the near future.  Classical fast approximations exist for some steps and in some cases, such as detector simulations for a particular experiment, but they are often not expressive enough to achieve high fidelity compared to a full simulation routine.

Deep neural network-based simulations (called \textit{deep generative models}) are a promising alternative to classical fast simulations.  Since the first deep generative model applied to high energy physics~\cite{deOliveira:2017pjk}, there have been a large number of proposals to use these tools for fast simulation and many other applications~\cite{hepmllivingreview,Butter:2022rso,Butter:2020tvl}.  In this paper, we revisit the original problem of emulating parton shower Monte Carlo simulations.  These simulations describe the formation of jets of hadrons that emerge from the high energy quarks and gluons.  Jets are ubiquitous at particle colliders and are the most complex objects reconstructed from hadronic final states. Together, these qualities make jets a standard benchmark for developing machine learning-based generative models.

Many deep generative models have been deployed to the problem of emulating jet formation.  The first approaches used images by spatially discretizing the radiation pattern within jets~\cite{Cogan:2014oua,deOliveira:2015xxd}.  Generative Adversarial Networks (GANs)~\cite{Goodfellow:2014upx,deOliveira:2017pjk} and autoencoders~\cite{Monk:2018zsb} were able to reproduce many aspects of the parton shower, but were fundamentally limited because of their pixelization.  While other applications of deep generative models naturally process image data (e.g. calorimeter simulations~\cite{Paganini:2017dwg,Paganini:2017hrr,Carminati:2018khv,Chekalina:2018hxi,Erdmann:2018jxd,deOliveira:2017rwa,Belayneh:2019vyx,Buhmann:2020pmy,2009.03796,Buhmann:2021caf,Rehm:2021zoz,Khattak:2021ndw,Krause:2021ilc,Krause:2021wez,ATLAS:2021pzo,ATLAS:2022jhk,Bieringer:2022cbs,Rogachev:2022hjg,Buhmann:2021lxj,AbhishekAbhishek:2022wby,Liu:2022dem,Diefenbacher:2023vsw,Krause:2022jna,Cresswell:2022tof,Mikuni:2022xry}), jets are naturally represented as variable-sized point clouds and information is lost when they are projected onto fixed size grids with reduced spatial position information compared to the original detector granularity.

Point cloud generative models (PCGM) offer the solution to the inherent challenges with pixelation.  The first PCGM applied to jet formation was Ref.~\cite{Andreassen:2018apy}, which used a recurrent model to describe the probability density of a given jet. Recently, there has been a surge of interest in more general PCGMs that do not need to make any assumptions or approximations about the underlying generative process.  This latest wave of methods began with a graph neural network-based GAN~\cite{Kansal:2021cqp} and now includes a deep sets-based GAN~\cite{Buhmann:2023pmh} and a normalizing flow~\cite{Kach:2022qnf,Verheyen:2022tov}.  These models mark a significant step forward in the application of generative models to particle physics, but there is still significant room for improvement in both precision and robustness.  For example, GANs solve a minimax problem and are thus difficult to train.  Normalizing flows are more stable to train, but may have difficulties when generating low level inputs (such as particle kinematic information), or a variable-length representation with complex topology due to the invertible nature of their neural networks.


\begin{figure*}[ht]
    \centering
    \includegraphics[width=.95\textwidth]{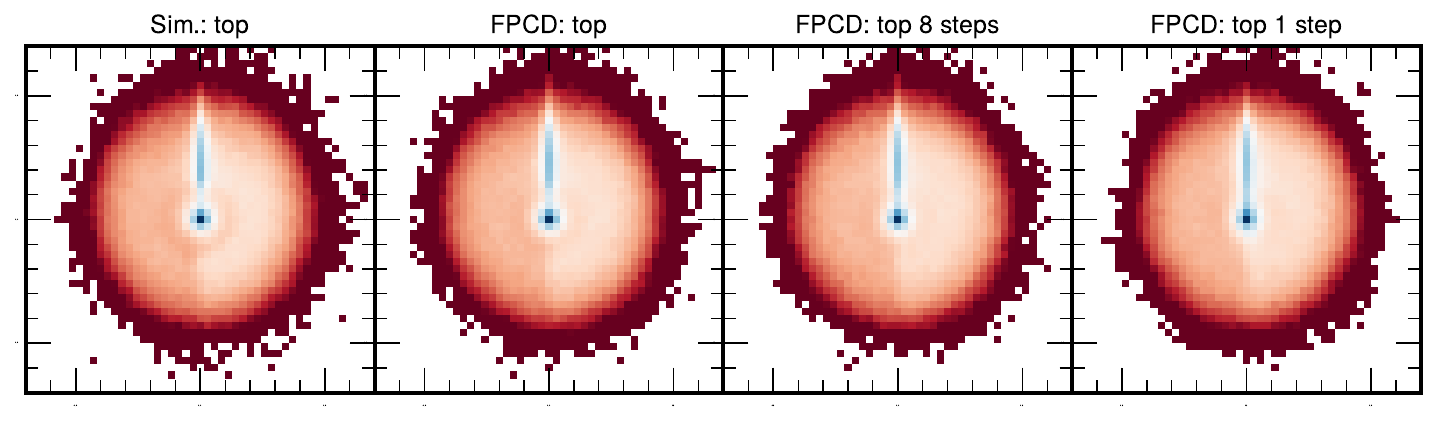}
    \caption{Average top quark initiated jet in the full simulation, after generation with the diffusion model, and after distillation resulting in 8 or a single time step used during sampling.}
    \label{fig:visual}
\end{figure*}

In the machine learning literature, the most precise generative neural networks are diffusion models (see e.g. Ref.~\cite{dhariwal2021diffusion}).  These approaches circumvent the challenges with other models by performing a convex optimization problem, but without the need for invertible transformations.  This can be achieved by learning the score of the probability density $\nabla \log p$ instead of the probability density directly.   The first diffusion model applied to particle physics was in the context of image-based calorimeter simulation~\cite{Mikuni:2022xry}, significantly extending the dimensionality of previous results.  Our goal is to adapt diffusion models to the variable-length point cloud setting for parton showers and other phenomena in high energy physics while also reducing the generation time to be competitive with other fast generation methods. To this end, we introduce our algorithm for Fast Point Cloud Generation (FPCD), used to simulate point cloud data with varying length much faster than a standard diffusion implementation. Examples of generated point clouds using our proposed algorithm are shown in Fig.~\ref{fig:visual}, where we compare the average energy deposition for top quark initiated jets generated by the full simulation or by the generative model. We accelerate the sampling time of the surrogate model using a method called progressive distillation~\cite{salimans2022progressive}, resulting in a generative model with high physics fidelity and fast sampling times.

While this paper was being finalized, the authors of Ref.~\cite{Leigh:2023toe} also proposed a diffusion-based PCGM for jet formation.  The proposal in our paper differs from Ref.~\cite{Leigh:2023toe} in a few ways. First, our model does not condition on the jet mass, but rather utilizes a separate diffusion model to determine the jet kinematics. Next, it is much faster (via progressive distillation) and is conditioned on the particle type, thereby avoiding the training of multiple diffusion models for each type of jet. Finally, we also provide results for more particle types (including gluons and $W$ and $Z$ bosons in addition to light and top quarks) in two different datasets with varying number of particles to demonstrate that our model is capable of generating outputs of varying sizes.

This paper is organized as follows.  Section~\ref{sec:method} introduces score-based diffusion models and describes how they can be accelerated with progressive distillation.  We then detail our implementation of the diffusion-based generative model for parton showers in Sec.~\ref{sec:dataset}.  Numerical results are presented in Sec.~\ref{sec:results} and the paper ends with conclusions and outlook in Sec.~\ref{sec:conclusions}.


\section{Score-based generative models and progressive distillation}
\label{sec:method}
The goal of a generative model is to be able to generate new observations from a noise distribution. Diffusion models became popular in recent years for their capacity to generate realistic data, often surpassing standard state-of-the-art generative models. In score-based methods~\cite{Song2021ScoreBasedGM}, a diffusion process is designed to slowly perturb the data through the addition of noise, while a neural network learns a time-dependent score function $\nabla_x\log p_\text{data}$ for some high-dimensional distribution $\mathbf{x}\in\mathbb{R}^\mathrm{D}$ described by the probability density $p_\text{data}$. The score function is then used in a reverse-diffusion process: starting from a noisy distribution and proceeding to denoise the observation.  The diffusion model is described by latent variables $\mathbf{z} = \left\{ \mathbf{z}_t| t \in [0,1]\right\}$ with a time-dependent noise schedule $\alpha_t, \sigma_t$, such that the log signal-to-noise-ratio $\log[\alpha^2_t/\sigma^2_t]$, decreases monotonically with time. 
During training, the network learns to denoise $\mathbf{z}_t\sim q(\mathbf{z}_t|\mathbf{x})=\mathcal{N}(\mathbf{z}_t;\alpha_t\mathbf{x},\sigma_t^2\mathbf{I})$ towards the unperturbed data $\mathbf{x}\sim p_\text{data}$, effectively learning an estimate $\hat{\mathbf{x}}_\theta \approx \mathbf{x}$ by updating the trainable parameters $\theta$ during training. Following Ref.~\cite{salimans2022progressive}, we instead train a network to estimate a ``velocity" parameter $\mathbf{v} \equiv \alpha_t\mathbf{\epsilon}-\sigma_t\mathbf{x}$, with $\mathbf{\epsilon}\sim \mathcal{N}(\mathbf{0},\mathbf{I})$ which is observed to yield accurate results while also simplifying the distillation method employed later. The loss function to be minimized during optimization is then defined as:
\begin{equation}
    \mathcal{L}_\theta = \mathbb{E}_{\mathbf{\epsilon},t} \left\| \mathbf{v}_t - \hat{\mathbf{v}}_{t,\theta}\right\|^2,
\end{equation}
where $t$ is sampled uniformly over the considered interval. 
In this formulation, we can identify the estimate of the score function as:
\begin{equation}
    \nabla_z\log \hat{p}_\theta(\mathbf{z}_t) =  \mathbf{z}_t  - \frac{\alpha_t}{\sigma_t}\hat{\mathbf{v}}_\theta(\mathbf{z}_t).
\end{equation}
In our implementation, we consider the variance-preserving setting of diffusion processes, where $\sigma_t^2 = 1- \alpha_t^2$. For the time-dependence, we use a cosine schedule such that $\alpha_t = \cos(0.5\pi t)$.

The generation of new samples is then carried out using the DDIM sampler proposed in Ref.~\cite{DBLP:journals/corr/abs-2010-02502} that uses an integration rule to solve the deterministic ordinary differential equation:
\begin{equation}
     \mathrm{d}\mathbf{z}_t = [f(\mathbf{z}_t,t)-\frac{1}{2}g^2(t)\nabla_z\log \hat{p}_\theta(\mathbf{z}_t)]\mathrm{d}t,
     \label{eq:int}
\end{equation}
with drift coefficient $f(\mathbf{z}_t,t) = \frac{\mathrm{d} \log\alpha_t}{\mathrm{d} t}\mathbf{z}_t$ and diffusion coefficient $g^2(t) = \frac{\mathrm{d} \sigma_t^2}{\mathrm{d} t}$. In the DDIM solver, the update rule is then specified by:
\begin{equation}
    \mathbf{z}_s = \alpha_s\hat{\mathbf{x}}_\theta(\mathbf{z}_t)  + \sigma_s\frac{\mathbf{z}_t -\alpha_t\hat{\mathbf{x}}_\theta(\mathbf{z}_t)}{\sigma_t}. 
\end{equation}
In practice, solving Eq.~\ref{eq:int} can be slow since the error introduced by the numerical integration is sensitive to the number of time steps chosen, often requiring hundreds to thousands of time steps and hence function evaluations of the trained model. 

To accelerate diffusion models, Ref.~\cite{salimans2022progressive} introduced a technique called \emph{progressive distillation}. Starting from a trained diffusion model, the goal of progressive distillation is to learn iteratively to halve the number of time steps required during  generation of new samples. In this setting, the trained diffusion model (``teacher'') is used to initialize a ``student'' model. During training, the goal is to have the student model learn how to denoise data $\mathbf{z}_t$ towards a target $\mathbf{\widetilde{x}}$, where $\mathbf{\widetilde{x}}$ does not represent the clean data ($\mathbf{x}$) anymore, but instead is  one that makes a single student DDIM step to match two teacher DDIM steps. This process is then repeated multiple times, with the student at the end of each iteration becoming the new teacher. In this work, we train a diffusion model with initial number of steps $N$ fixed to 512. From there, we distill the model multiple times, reporting the results obtained with $N = 512$, $N=8$, and $N=1$.


\section{Point Cloud Diffusion for Collider Data}
\label{sec:dataset}

We train a conditional diffusion model to generate particle jets conditioned on the initial particle type. We use the datasets introduced in Ref.~\cite{Kansal:2021cqp} consisting of jets initiated by light-quarks, gluons, top quarks, $W$ and $Z$ bosons. The jets are generated with transverse momenta $\pt$ around 1~TeV and are clustered using the anti-$k_t$ algorithm~\cite{Cacciari:2008gp} with a radius parameter of 0.8. Each jet has a maximum number of particles stored fixed to 30~\cite{kansal_raghav_2022_6975118} or 150~\cite{kansal_raghav_2022_6975117}. For each jet, the four-momentum information $(\pt_{jet},\eta_{jet},\phi_{jet}, m_{jet})$ is provided, as well as the particle multiplicity. For each particle clustered inside a jet, the relative set of kinematic quantities are provided:
\begin{equation}
\begin{split}    
    \pt_{rel} &= \pt_{part}/\pt_{jet} \\
    \eta_{rel} &= \eta_{jet}- \eta_{part}\\
    \phi_{rel} &= \phi_{jet}- \phi_{part}.
\end{split}    
\end{equation}

Our goal is to develop a diffusion model that is conditioned on the particle's type and is able to generate both jet- and particle-level kinematic information. To accomplish this task, we train two diffusion models simultaneously. The first model learns the jet kinematic information, including particle multiplicity, while the second is conditioned on the jet kinematic distributions to generate particle information. Effectively, the loss function that is minimized during training is 
\begin{equation}
    \mathcal{L}_{\mathrm{jet},\mathrm{particle}} = \mathcal{L}_{\mathrm{jet}} + \mathcal{L}_{\mathrm{particle}},
\end{equation}
with different models trained to generate  jet information and  particle information. 
During the generation step, we first sample the jet kinematic information together with the particle multiplicity, conditioned on the type of the jet we aim to generate. This information is then used as an input to generate the particle information for each jet. The particle multiplicity generated determines the total number of particles generated in each jet. Although it is feasible to achieve a perfect match between the output particle multiplicity and the sampled particle multiplicity, we prefer to employ a masking plus zero-padding approach. This involves always sampling a set number of particles (either 30 or 150 depending on the dataset), but in the generation process, we mask the input noise and only consider the desired particle multiplicity.

Prior to training, the inputs to the diffusion model  undergo a normalization process where all input features are standardized by adjusting their mean and standard deviation to zero and one, respectively. 

\begin{figure*}[]
    \centering
    \includegraphics[width=\textwidth]{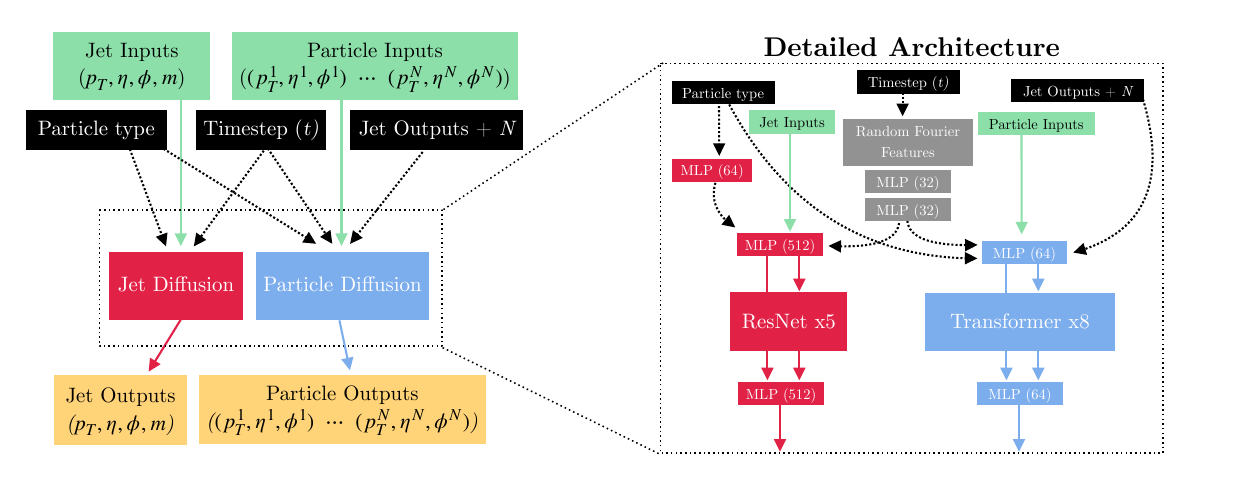}
    \caption{Description of the network architectures used to train the jet and particle diffusion models. Numbers after layers represent the number of hidden nodes associated to the layer. See the text for more information. }
    \label{fig:architecture}
\end{figure*}


The generative model designed to produce jet kinematic information is based on a fully-connected architecture incorporating multiple skip connections. Specifically, the model employs five \textsc{ResNet}~\cite{he2016deep} blocks, where each residual layer is connected to the output of a two-layer network through a skip connection. The activation function used is \textsc{LeakyRelu}~\cite{DBLP:journals/corr/XuWCL15} with a slope of $\alpha=0.01$, and all layer sizes are set to 512.

The particle diffusion model employs a \textsc{DeepSets}~\cite{DBLP:journals/corr/ZaheerKRPSS17} architecture with Transformer layers ~\cite{DBLP:journals/corr/VaswaniSPUJGKP17} to increase the model's expressivity. The input sets are first mapped into a larger latent space using a fully-connected layer with a size of 64, applied independently to each particle in the set. The model then employs eight Transformer encoding blocks followed by a fully-connected layer with a size of 64 before the output layer. The activation function used is again \textsc{LeakyRelu}, and the outputs of the transformer layers are summed to the last layer before the first Transformer block, which is observed to result in better performance according to our experiments.

Both diffusion models incorporate time information by feeding random Fourier features \cite{tancik2020fourier} through two fully connected layers with 32 and 64 nodes. The resulting embeddings are combined with additional conditional information including jet type for the jet diffusion model and both jet type and jet kinematic information for the particle diffusion model. After passing through a fully connected layer of size 64, these embeddings are concatenated with the inputs of each diffusion model.

A visual description of both models is shown in Fig.~\ref{fig:architecture}. 

The implementation of the model is carried out using \textsc{Keras} backend~\cite{keras} with a \textsc{TensorFlow}~\cite{tensorflow} backend. The model is trained for up to 250 epochs with a cosine learning rate schedule~\cite{DBLP:journals/corr/LoshchilovH16a} with initial learning rate of $16\times10^{-4}$. If the loss function does not decrease for 20 consecutive epochs, evaluated in a separate testing set, representing 20\% to the sample size, the training is stopped.  During training, 16 NVIDIA A100 GPUs are used simultaneously interfaced with the Horovod package~\cite{sergeev2018horovod} on the Perlmutter supercomputer~\cite{Perlmutter}. The batch size in each GPU is set to 128. The hyperparameters used in the model architecture were optimized using the \textsc{KerasTuner}~\cite{omalley2019kerastuner} package with Hyperband~\cite{JMLR:v18:16-558} algorithm.

\begin{figure*}[ht]
    \centering
        \includegraphics[width=.35\textwidth]{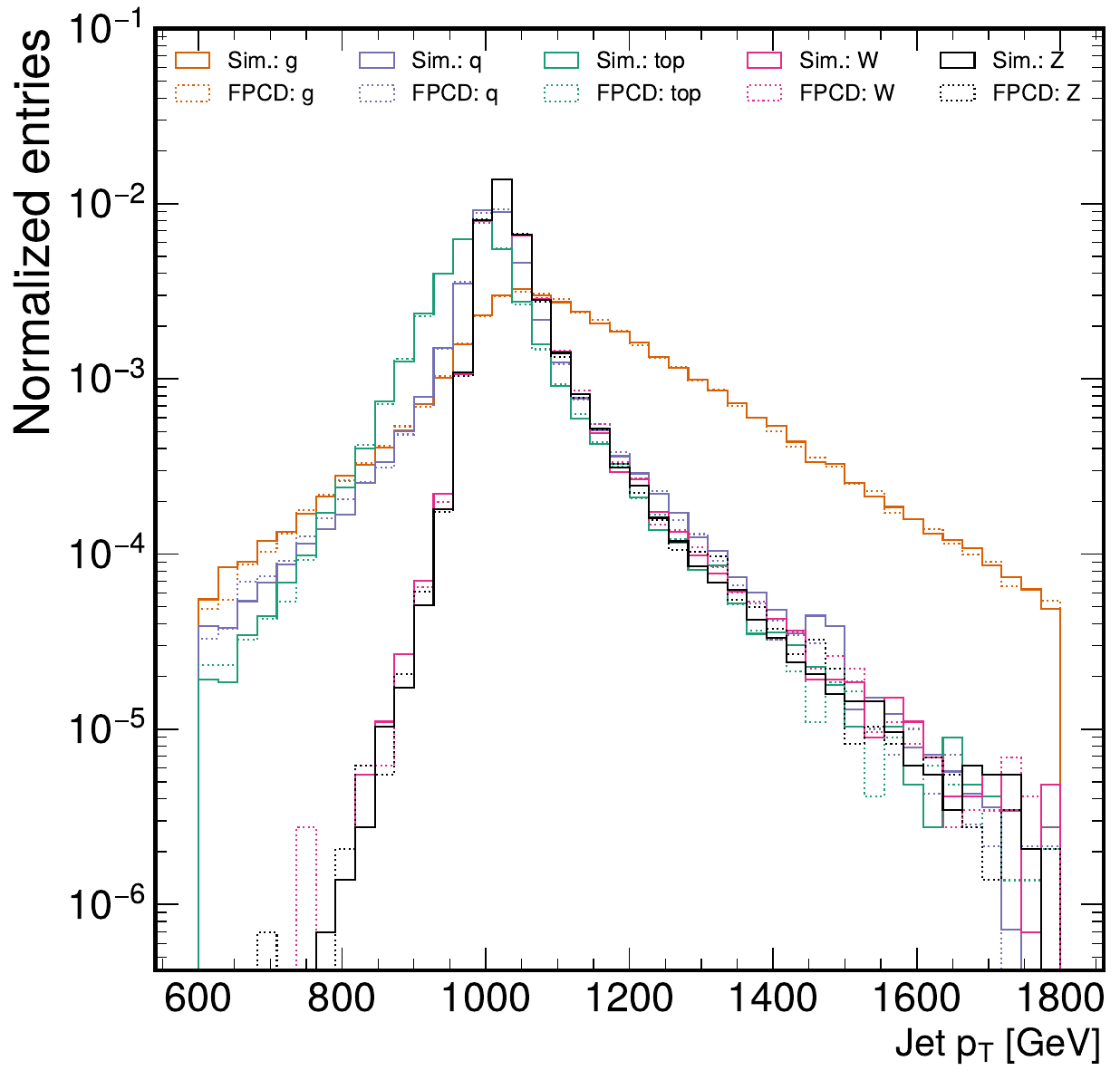}
        \includegraphics[width=.35\textwidth]{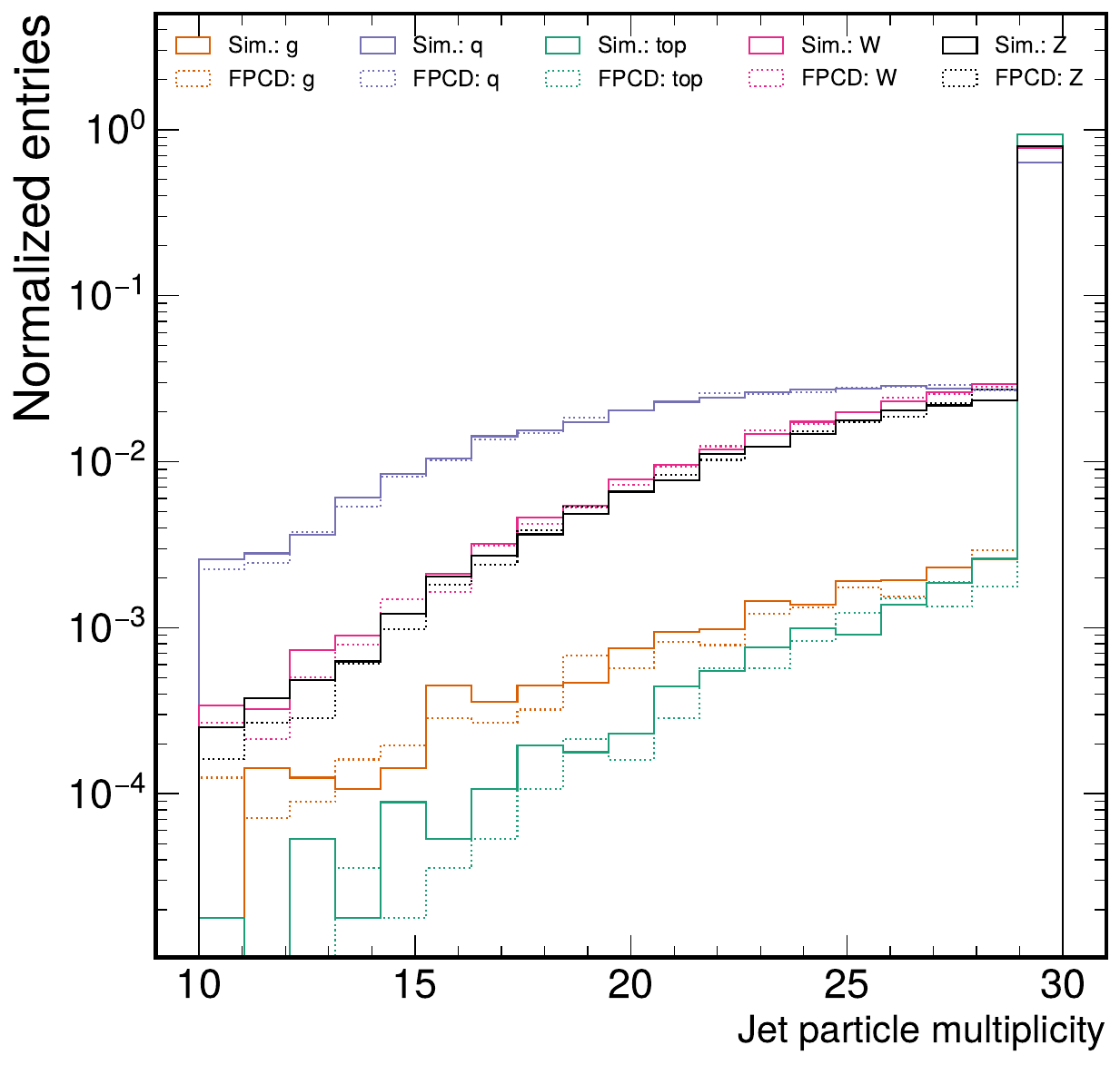} 
        \includegraphics[width=.35\textwidth]{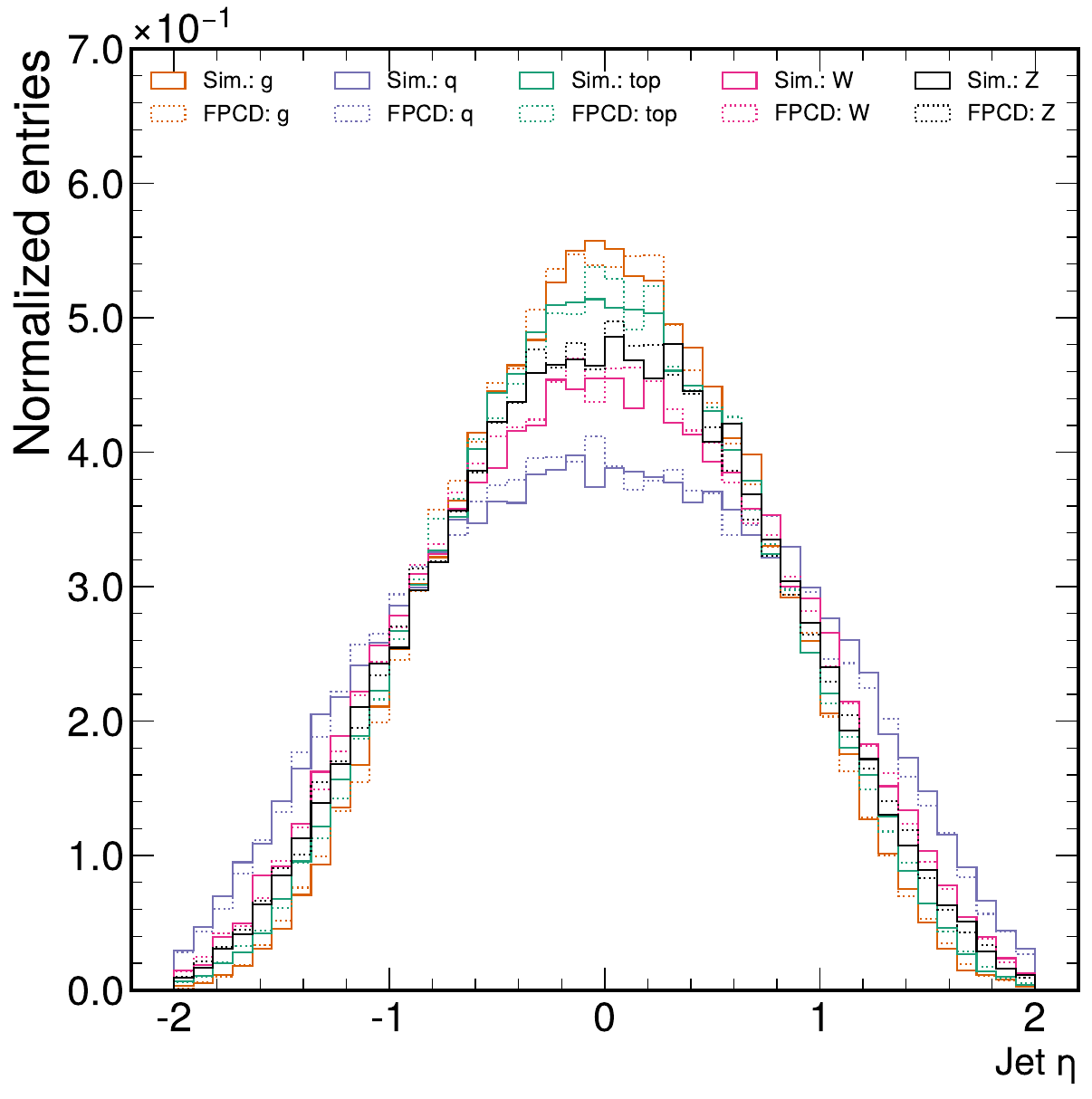}
        \includegraphics[width=.35\textwidth]{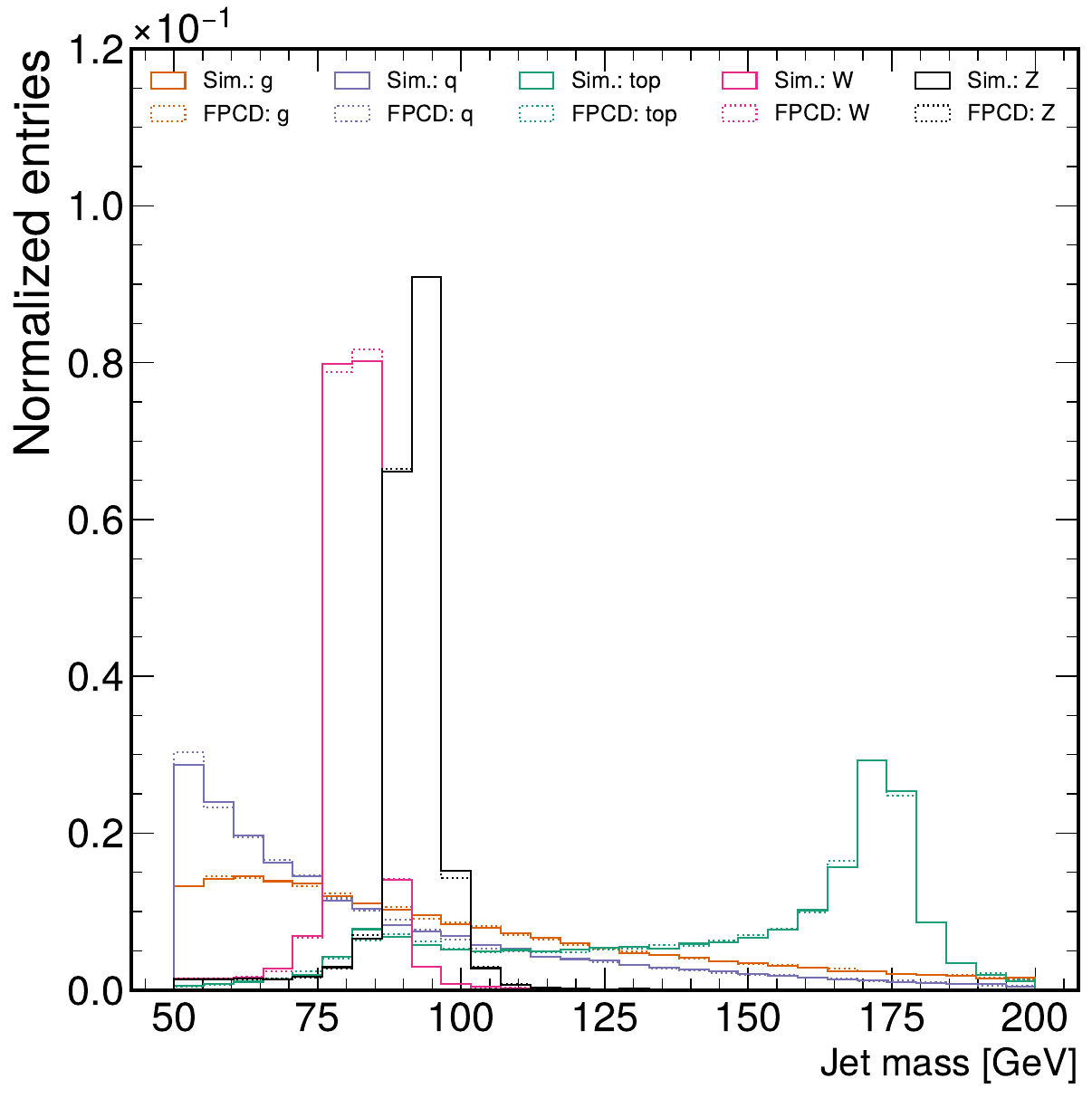}                
    \caption{Generated jet kinematic information using FPCD compared to simulated events for particle jets consisting of light-quarks (q), gluons (g), and top quarks (top).}
    \label{fig:jet}
\end{figure*}

\begin{figure*}[ht]
    \centering
        \includegraphics[width=.28\textwidth]{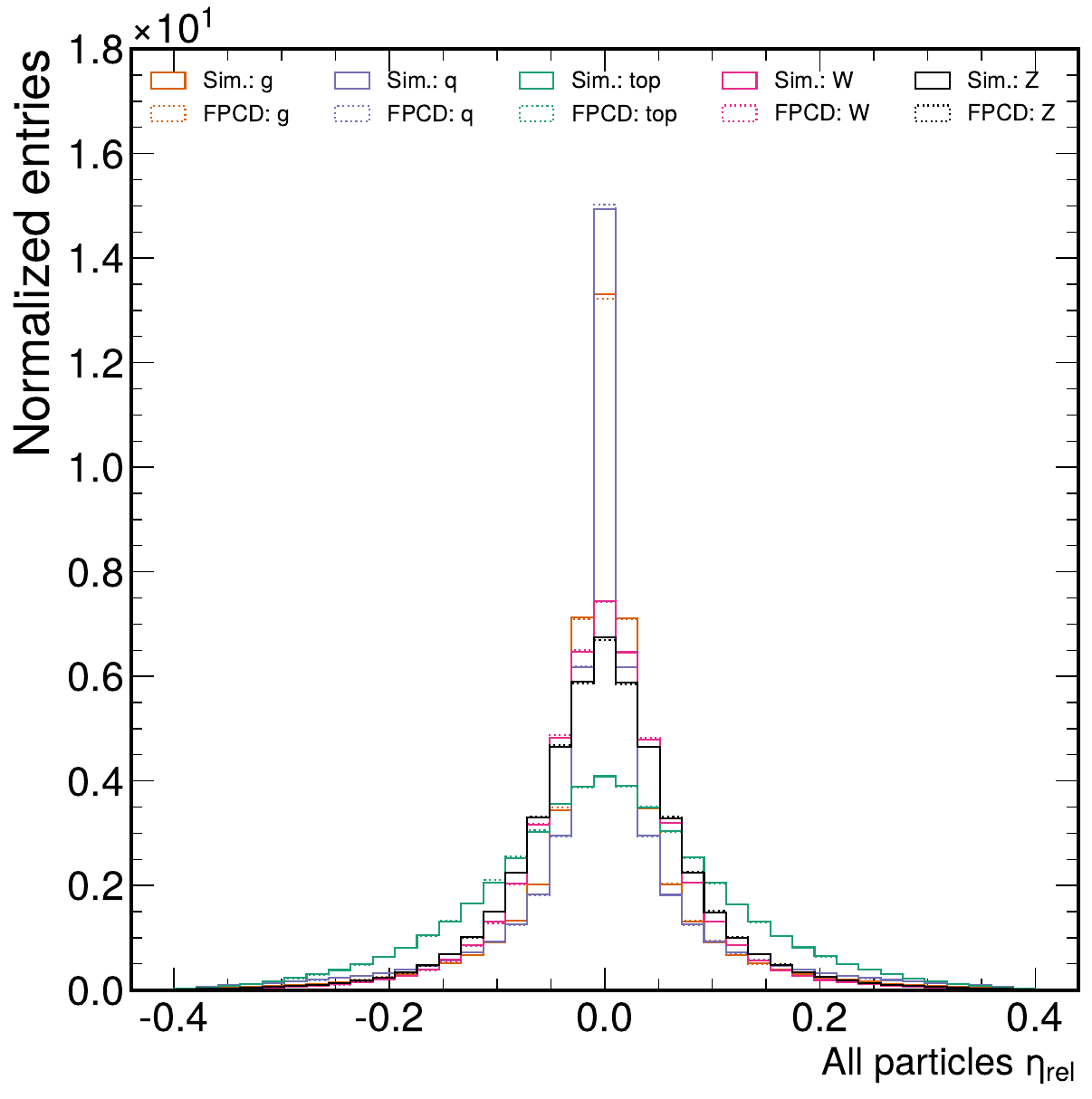}
        \includegraphics[width=.28\textwidth]{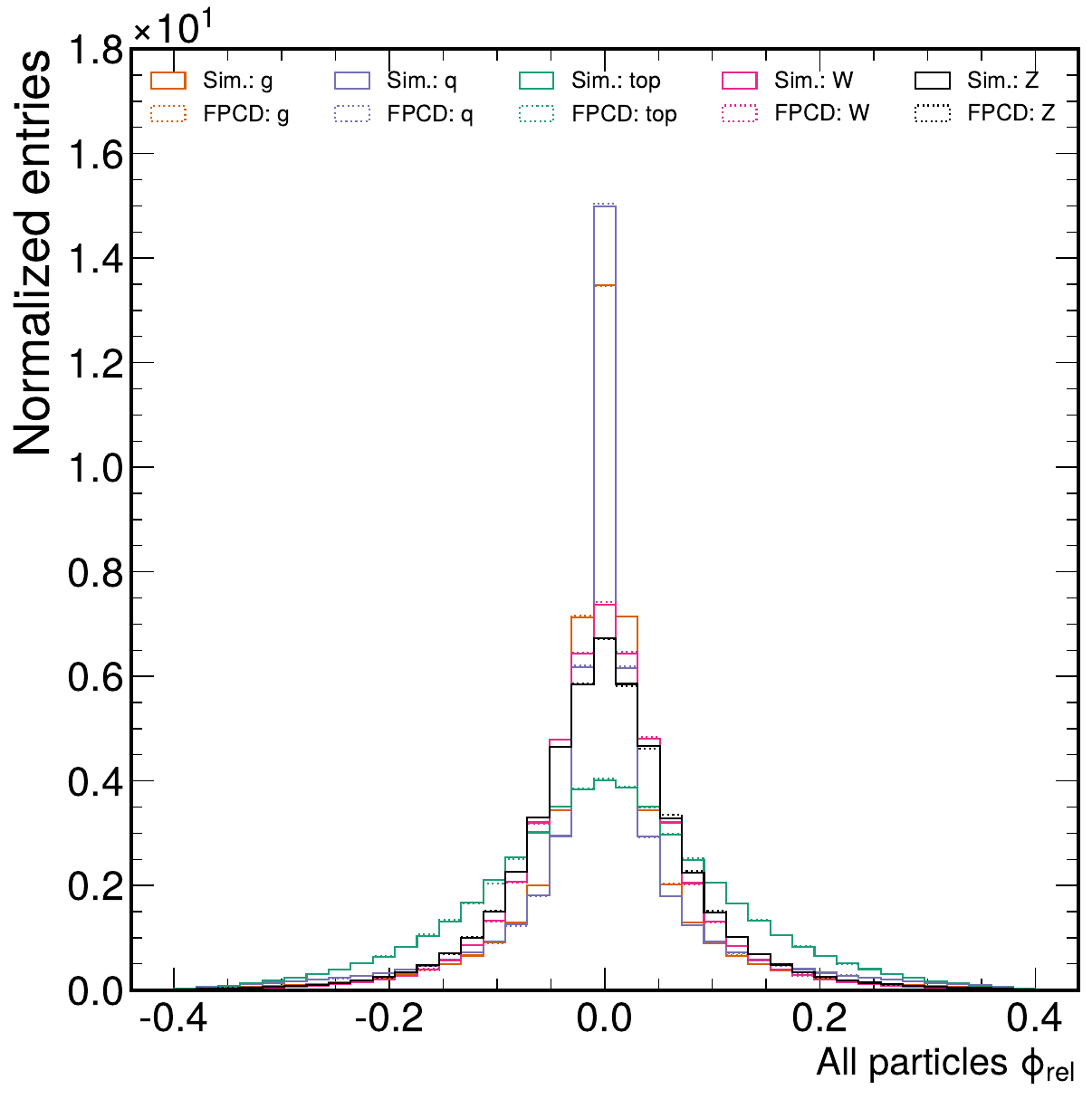}
        \includegraphics[width=.28\textwidth]{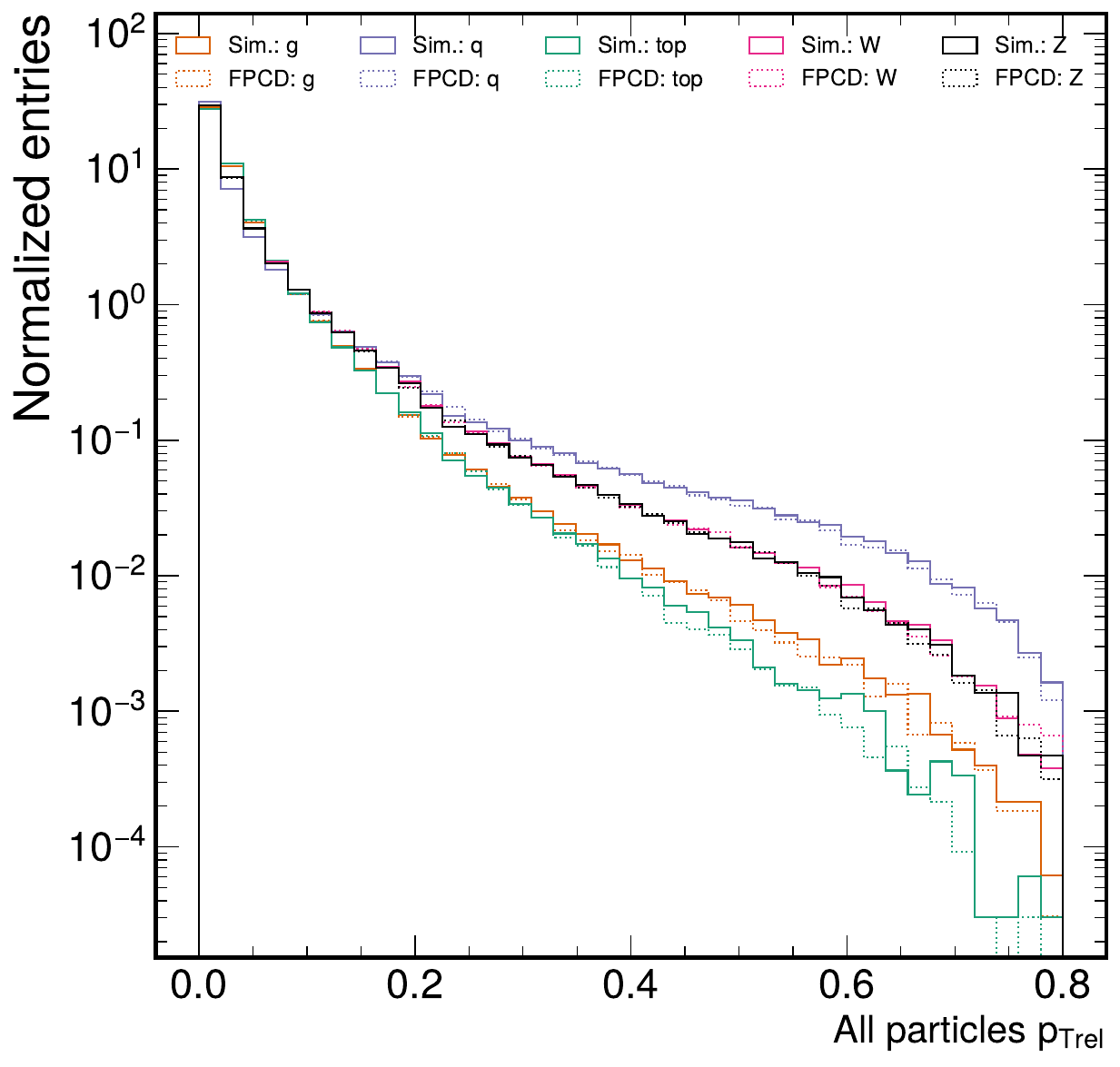}        
    \caption{Generated particle kinematic information using FPCD compared to simulated events for particle jets consisting of light-quarks (q), gluons (g), and top quarks (top). For each jet, all particles for both simulation and FPCD are shown.}
    \label{fig:part}
\end{figure*}

\section{Results}
\label{sec:results}
The performance of the generative model is evaluated using physics-based metrics proposed in~\cite{Kansal:2021cqp} as well as additional metrics designed specifically to assess the quality of the jet kinematic generation. These metrics include the 1-Wasserstein (W$_1$) distances that are calculated using only particle information such as averaged particle relative momentum W$^\mathrm{P}_1$, relative jet mass W$^\mathrm{PM}_1$, and average of first five energy flow polynomials W$^\mathrm{PEFP}_1$~\cite{Komiske:2017aww}. The 1-Wasserstein distances are also calculated for jet kinematic information,  including jet transverse momentum W$^\mathrm{JP}_1$, jet pseudorapidity W$^\mathrm{J\eta}_1$, jet mass W$^\mathrm{JM}_1$, and jet particle multiplicity W$^\mathrm{JN}_1$. The evaluation also includes Fréchet ParticleNet distance (FPND), coverage (Cov), and minimum matching distance (MMD), described in Ref.~\cite{Kansal:2021cqp}. To calculate each metric, 50,000 generated examples for each jet category are compared against 50,000 validation samples that were not used during training. Uncertainties are estimated using bootstrapping with replacement following~\cite{Kansal:2021cqp}. Additionally, results for distilled models with different number of total time steps are also provided.

Different jet kinematic distributions are shown in Fig.~\ref{fig:jet} as well as the comparison of the different metrics listed in Tab.~\ref{tab:results_jet}. We also present the per-particle distributions in Fig.~\ref{fig:part}, displaying simultaneously all particles inside a given jet.  Results are also compared with the official implementations of \textsc{EPiC-GAN} and \textsc{MP-GAN}, where in the latter the MP-MP implementation is taken for the comparison.

 While other implementations using the same datasets exist~\cite{Leigh:2023toe,Kach:2022qnf}, these models are not directly comparable to our method as they are conditioned by the jet kinematic information, whereas our method simultaneously models both the jet and particle kinematic distriburions.

\begin{table*}[htb]
\centering
\caption{Comparison of the results obtained between different generative models in the task of particle property generation in the dataset consisting of 30 particles. Baseline FPCD uses 512 time steps during sampling. Distilled models are listed alongside number of time steps used. Lower is better for all metrics except Cov. FPND metrics are not available for W and Z bosons, hence omitted.}
\label{tab:results_part}
\begin{tabular}{ | c | c | c | c | c | c | c | c | c | c | c | c | c |}
\hline
Jet class & Model & W$^\mathrm{PM}_1$ ($\times 10^{-3}$) & W$^\mathrm{P}_1$ ($\times 10^{-3}$) & W$^\mathrm{PEFP}_1$ ($\times 10^{-5}$) & FPND & Cov$\uparrow$ & MMD \\
\hline
          & FPCD & \textbf{0.36 $\pm$ 0.08} & \textbf{0.34 $\pm$ 0.09} & \textbf{0.47 $\pm$ 0.13 }& \textbf{0.07} & \textbf{0.55} & \textbf{0.03} \\
          & FPCD 8 & 0.60 $\pm$ 0.16 & \textbf{0.36 $\pm$ 0.07} & \textbf{0.54 $\pm$ 0.09} & \textbf{0.07} & \textbf{0.55} & \textbf{0.03} \\
    Gluon & FPCD 1 & 0.65 $\pm$ 0.11 & \textbf{0.34 $\pm$ 0.06} & 0.60 $\pm$ 0.09 & 0.11 & \textbf{0.55} & \textbf{0.03} \\
          & \textsc{MP-GAN}~\cite{Kansal:2021cqp} & 0.69 $\pm$ 0.07 & 1.8 $\pm$ 0.2 &0.9 $\pm$ 0.6 &0.20 &0.54 &0.037 \\
          & \textsc{EPiC-GAN}~\cite{Buhmann:2023pmh} & \textbf{0.3 $\pm$ 0.1} & 1.6 $\pm$ 0.2& \textbf{ }\textbf{0.4 $\pm$ 0.2} & 1.01 $\pm$ 0.07& -& -\\
\hline
          & FPCD &  \textbf{0.52 $\pm$ 0.07} & \textbf{0.27 $\pm$ 0.06} & \textbf{0.38 $\pm$ 0.11} & \textbf{0.08} & 0.49 & \textbf{0.02} \\
          & FPCD 8 & \textbf{0.59 $\pm$ 0.14} & 0.35 $\pm$ 0.05 & \textbf{0.44 $\pm$ 0.07} & 0.09 & 0.48 & \textbf{0.02} \\
Light Quark& FPCD 1 & \textbf{0.59 $\pm$ 0.08} & 0.36 $\pm$ 0.08 & 0.50 $\pm$ 0.08 & 0.09 & 0.48 & \textbf{0.02} \\
          & \textsc{MP-GAN}~\cite{Kansal:2021cqp} & \textbf{0.6 $\pm$ 0.2} & 4.9 $\pm$ 0.5 & 0.7 $\pm$ 0.4 & 0.35 &  \textbf{0.50} &  0.026\\
          & \textsc{EPiC-GAN}~\cite{Buhmann:2023pmh} & \textbf{0.5 $\pm$ 0.1} & 4.0 $\pm$ 0.4 & 0.8 $\pm$ 0.4 & 0.43 $\pm$ 0.03 & -& -\\
\hline
          & FPCD & \textbf{0.51 $\pm$ 0.07} & \textbf{0.41 $\pm$ 0.12} & \textbf{1.25 $\pm$ 0.19} & \textbf{0.17} & \textbf{0.58} & \textbf{0.05} \\
          & FPCD 8 & 0.80 $\pm$ 0.06 & \textbf{0.45 $\pm$ 0.12} & 1.91 $\pm$ 0.30 & 0.37 & \textbf{0.58} & \textbf{0.05} \\
Top Quark & FPCD 1 & 1.22 $\pm$ 0.09 &\textbf{ 0.46 $\pm$ 0.10} & 2.66 $\pm$ 0.26 & 0.56 & 0.57 & \textbf{0.05} \\
          & \textsc{MP-GAN}~\cite{Kansal:2021cqp} & \textbf{0.6 $\pm$ 0.2} & 2.3 $\pm$ 0.3 & 2 $\pm$ 1 & 0.37 & 0.57 & 0.071 \\
          & \textsc{EPiC-GAN}~\cite{Buhmann:2023pmh} & \textbf{0.5 $\pm$ 0.1} & 2.1 $\pm$  0.1 & 1.7 $\pm$ 0.3 & 0.31 $\pm$ 0.037 & -& -\\
\hline
          & FPCD & \textbf{0.26 $\pm$ 0.03} & \textbf{0.39 $\pm$ 0.08} & \textbf{0.15 $\pm$ 0.02} & - & \textbf{0.56} & \textbf{0.02} \\
W Boson   & FPCD 8 & 0.48 $\pm$ 0.04 & \textbf{0.38 $\pm$ 0.05} & 0.22 $\pm$ 0.02 & - & 0.55 & \textbf{0.02} \\
          & FPCD 1 & 0.94 $\pm$ 0.06 & \textbf{0.42 $\pm$ 0.09} & 0.35 $\pm$ 0.03 & - & \textbf{0.56} & \textbf{0.02} \\

\hline
          &  FPCD & \textbf{0.21 $\pm$ 0.04} & \textbf{0.40 $\pm$ 0.13} & \textbf{0.18 $\pm$ 0.03} & - &  \textbf{0.56} & \textbf{0.02} \\
Z Boson   & FPCD 8 & 0.40 $\pm$ 0.04 & \textbf{0.35 $\pm$ 0.04} & 0.27 $\pm$ 0.03 & - &  \textbf{0.56} & \textbf{0.02} \\
          & FPCD 1 & 0.99 $\pm$ 0.05 & \textbf{0.35 $\pm$ 0.06} & 0.49 $\pm$ 0.03 &- & \textbf{0.56} & \textbf{0.02} \\
\hline

\end{tabular}
\end{table*}

\begin{table*}[htb]
\centering
\caption{Comparison of the results obtained between different generative models for the task of jet property generation in the dataset consisting of 30 particles. Baseline FPCD uses 512 time steps during sampling. Distilled models are listed alongside number of time steps used. Lower is better for all metrics listed except Cov. }
\label{tab:results_jet}
\begin{tabular}{ | c | c | c | c | c | c | c | c | c | c | c | c | c |}
\hline
Jet class & Model &  W$^\mathrm{J\pt}_1$ & W$^\mathrm{J\eta}_1$& W$^\mathrm{JM}_1$ &W$^\mathrm{JN}_1$ \\
\hline
          & FPCD & 1.5 $\pm$ 0.5 & 0.009 $\pm$ 0.003 & 0.30 $\pm$ 0.07 & 0.020 $\pm$ 0.009\\
Gluon     & FPCD 8 & 1.5 $\pm$ 0.5 & 0.010 $\pm$ 0.003 & 0.30 $\pm$ 0.07 & 0.021 $\pm$ 0.009 \\
          & FPCD 1 & 1.5 $\pm$ 0.4 & 0.011 $\pm$ 0.003 & 0.30 $\pm$ 0.10 & 0.025 $\pm$ 0.011\\

\hline
          &  FPCD & 1.3 $\pm$ 0.4 & 0.008 $\pm$ 0.002 & 0.39 $\pm$ 0.14 & 0.023 $\pm$ 0.009\\
Light Quark  & FPCD 8 & 1.4 $\pm$ 0.3 & 0.009 $\pm$ 0.002 & 0.39 $\pm$ 0.14 & 0.024 $\pm$ 0.010 \\
          & FPCD 1 &  1.4 $\pm$ 0.3 & 0.009 $\pm$ 0.002 & 0.39 $\pm$ 0.09 & 0.024 $\pm$ 0.008\\

\hline
          & FPCD & 1.4 $\pm$ 0.3 & 0.009 $\pm$ 0.003 & 0.37 $\pm$ 0.12 & 0.022 $\pm$ 0.007\\
Top Quark & FPCD 8 & 1.5 $\pm$ 0.3 & 0.010 $\pm$ 0.003 & 0.37 $\pm$ 0.12 & 0.023 $\pm$ 0.007 \\
          & FPCD 1 & 1.4 $\pm$ 0.3 & 0.009 $\pm$ 0.002 & 0.41 $\pm$ 0.12 & 0.025 $\pm$ 0.009\\
\hline
          & FPCD & 1.19 $\pm$ 0.34 & 0.008 $\pm$ 0.004 & 0.33 $\pm$ 0.15 & 0.021 $\pm$ 0.010 \\
W Boson   & FPCD 8 &1.23 $\pm$ 0.33 & 0.009 $\pm$ 0.003 & 0.34 $\pm$ 0.14 & 0.021 $\pm$ 0.010 \\
          & FPCD 1 & 1.21 $\pm$ 0.25 & 0.009 $\pm$ 0.003 & 0.33 $\pm$ 0.10 & 0.023 $\pm$ 0.011\\
\hline
          & FPCD & 1.14 $\pm$ 0.22 & 0.011 $\pm$ 0.004 & 0.34 $\pm$ 0.18 & 0.023 $\pm$ 0.013  \\
Z Boson   & FPCD 8 & 1.18 $\pm$ 0.24 & 0.012 $\pm$ 0.004 & 0.35 $\pm$ 0.18 & 0.024 $\pm$ 0.013\\
          & FPCD 1 & 1.43 $\pm$ 0.35 & 0.010 $\pm$ 0.004 & 0.36 $\pm$ 0.13 & 0.030 $\pm$ 0.015 \\
\hline  
\end{tabular}
\end{table*}

Similarly, we also consider the same physics-inspired metrics to evaluate FPCD in the dataset consisting of up to 150 particles per jet. In this case, the majority of the jets used during training need to be zero-padded and is used to display the capability of FPCD to learn how to generate jets with varying number of particles. The comparison of the physics inspired metrics are listed in Tab.~\ref{tab:results_part_150}. Since the jet kinematic information is not affected by the maximum number of particles stored in each jet, we only report the W$^\mathrm{JN}_1$ metric for each dataset in Tab.~\ref{tab:results_jet_150}. Histograms for each of the distributions considered in this study are provided in Appendix~\ref{app:results_150}.

\begin{table*}[htb]
\centering
\caption{Comparison of the results obtained between different generative models in the task of particle property generation in the dataset consisting of 150 particles. Baseline FPCD uses 512 time steps during sampling. Distilled models are listed alongside number of time steps used. Lower is better for all metrics except Cov. }
\label{tab:results_part_150}
\begin{tabular}{ | c | c | c | c | c | c | c | c | c | c | c | c | c |}
\hline
Jet class & Model & W$^\mathrm{PM}_1$ ($\times 10^{-3}$) & W$^\mathrm{P}_1$ ($\times 10^{-3}$) & W$^\mathrm{PEFP}_1$ ($\times 10^{-5}$) & Cov$\uparrow$ & MMD \\
\hline
          & FPCD & \textbf{0.44 $\pm$ 0.11} & \textbf{0.28 $\pm$ 0.05} & \textbf{0.91 $\pm$ 0.16} & \textbf{0.56} & \textbf{0.03} \\
          & FPCD 8 & 0.56 $\pm$ 0.06 & 0.40 $\pm$ 0.05 & \textbf{1.09 $\pm$ 0.23} & \textbf{0.56} & \textbf{0.03} \\
    Gluon & FPCD 1 & 0.65 $\pm$ 0.12 & 0.58 $\pm$ 0.03 & 1.49 $\pm$ 0.34 & 0.55 & \textbf{0.03} \\
          & \textsc{EPiC-GAN}~\cite{Buhmann:2023pmh} & \textbf{0.4 $\pm$ 0.1} & 3.2 $\pm$ 0.2 & \textbf{1.1 $\pm$ 0.7} & -& -\\
\hline
          & FPCD &  \textbf{0.46 $\pm$ 0.05} & \textbf{0.24 $\pm$ 0.02} & \textbf{0.43 $\pm$ 0.09} & \textbf{0.54} & \textbf{0.02} \\
          & FPCD 8 & \textbf{0.46 $\pm$ 0.09} & 0.39 $\pm$ 0.02 & 0.63 $\pm$ 0.21 & 0.53 & 0.02 \\
Light Quark& FPCD 1 & \textbf{0.39 $\pm$ 0.04} & 0.61 $\pm$ 0.03 & 0.57 $\pm$ 0.10 & \textbf{0.54} & \textbf{0.02} \\
          & \textsc{EPiC-GAN}~\cite{Buhmann:2023pmh} & \textbf{0.4 $\pm$ 0.1} & 3.9 $\pm$ 0.3 & \textbf{0.7 $\pm$ 0.4} & -& -\\
\hline
          & FPCD & \textbf{0.40 $\pm$ 0.07} & \textbf{0.30 $\pm$ 0.03} & \textbf{2.23 $\pm$ 0.16} & \textbf{0.58} & \textbf{0.05} \\
          & FPCD 8 & 0.56 $\pm$ 0.08 & 0.56 $\pm$ 0.04 & 3.29 $\pm$ 0.11 & \textbf{0.58} & \textbf{0.05} \\
Top Quark & FPCD 1 & 0.85 $\pm$ 0.09 & 0.87 $\pm$ 0.03 & 3.82 $\pm$ 0.24 & \textbf{0.58} & \textbf{0.05} \\
          & \textsc{EPiC-GAN}~\cite{Buhmann:2023pmh} & 0.6 $\pm$ 0.1 & 3.7 $\pm$ 0.3 & \textbf{2.8 $\pm$ 0.7} & -& -\\
\hline
          & FPCD & \textbf{0.29 $\pm$ 0.02} &\textbf{ 0.23 $\pm$ 0.02} & \textbf{0.22 $\pm$ 0.04} & 0.55 & \textbf{0.02} \\
W Boson   & FPCD 8 & 0.47 $\pm$ 0.03 & 0.39 $\pm$ 0.01 & 0.31 $\pm$ 0.04 & \textbf{0.56} & \textbf{0.02} \\
          & FPCD 1 & 0.93 $\pm$ 0.04 & 0.67 $\pm$ 0.01 & 0.37 $\pm$ 0.03 & \textbf{0.56} & \textbf{0.02} \\

\hline
          & FPCD & \textbf{0.28 $\pm$ 0.05} & \textbf{0.22 $\pm$ 0.03} & \textbf{0.23 $\pm$ 0.03} & 0.55 & \textbf{0.02} \\
Z Boson   & FPCD 8 & 0.52 $\pm$ 0.04 & 0.42 $\pm$ 0.01 & 0.37 $\pm$ 0.05 & 0.56 & \textbf{0.02} \\
          & FPCD 1 & 1.04 $\pm$ 0.08 & 0.69 $\pm$ 0.02 & 0.62 $\pm$ 0.06 & \textbf{0.57} & \textbf{0.02} \\
\hline

\end{tabular}
\end{table*}

\begin{table*}[htb]
\centering
\caption{Comparison of the results obtained between different generative models for the task of jet particle multiplicity generation in the dataset consisting of 150 particles. Baseline FPCD uses 512 time steps during sampling. Distilled models are listed alongside number of time steps used. }
\label{tab:results_jet_150}
\begin{tabular}{ | c | c | c | c | c | c | c | c | c | c | c | c | c |}
\hline
 & Model & Gluon & Light Quark & Top Quark & W Boson & Z Boson \\ 
\hline
          & FPCD & 0.157 $\pm$ 0.036 & 0.191 $\pm$ 0.067 & 0.171 $\pm$ 0.054 & 0.165 $\pm$ 0.056 &  0.241 $\pm$ 0.090 \\
 W$^\mathrm{JN}_1$  & FPCD 8 & 0.157 $\pm$ 0.035 & 0.190 $\pm$ 0.066 & 0.168 $\pm$ 0.054 & 0.165 $\pm$ 0.055 & 0.239 $\pm$ 0.090 \\
          & FPCD 1 & 0.157 $\pm$ 0.035 & 0.190 $\pm$ 0.067 & 0.169 $\pm$ 0.054 & 0.166 $\pm$ 0.055 & 0.239 $\pm$ 0.090 \\

\hline  
\end{tabular}
\end{table*}

Finally, we also compare the generation time for FPCD in Tab.~\ref{tab:time}

\begin{table}[htb]
\centering
\caption{Timing comparison for full jet generation with FPCD. A fixed batch size of 10'000 examples is used. The time reported is the sum of the time used to generate each jet and particle kinematic information in a single GPU. Full simulation time is taken from~\cite{Kansal:2021cqp}.}
\label{tab:time}
\begin{tabular}{ | c | c | c | c | c |}
\hline
  Model & 30 particles ($\mu$s) & 150 particles ($\mu$s)\\ 
\hline
        FPCD & 5$\times 10^3$ & 31$\times 10^3$\\
        FPCD 8 & 85 & 522\\
        FPCD 1 & 11 & 66 \\
        \textsc{EPiC-GAN}~\cite{Buhmann:2023pmh} & 2 & 12 \\
        \textsc{MP-GAN}~\cite{Kansal:2021cqp} \footnote{The \textsc{EPiC-GAN} work also provides a time comparison with a retrained MP-GAN. However since we do not retrain any model used for comparison, we decided to mention only the official results released in the original publications.} & 35.7 & - \\
        Full Simulation & 46$\times 10 ^3$ & 46$\times 10 ^3$\\
\hline  
\end{tabular}
\end{table}

The original FPCD model is highly accurate but computationally expensive. However, we found that a distilled model with as few as 8 time steps can achieve similar results while significantly reducing the overall sampling time. Surprisingly, a distilled model with only a single time step during generation still retains high fidelity while further reducing the sampling time. In Appendix~\ref{app:comp}, we compare the histograms for each of the kinematic distributions generated by the diffusion model and the distilled models in the dataset of top quark initiated jets.

\textsc{EPiC-GAN} is shown to be around one order of magnitude faster than FPCD even with a single diffusion step due to a different network architecture proposed by the authors. On the other hand, FPCD with a single time step is faster than \textsc{MP-GAN}, which also generates new jets through a single evaluation of the trained network. Nevertheless, all generative models are several orders of magnitude faster than the original physics simulation.

\section{Conclusion and Outlook}
\label{sec:conclusions}

In this work we introduced a fast point cloud diffusion (FPCD) model, providing flexibility, accuracy, and computational efficiency in jet generation. By simultaneously learning and generating multiple jet species, the two-part diffusion model generates both the jet kinematic information and particle information, conditioned on the jet kinematics and particle type to be generated.

Our model has achieved state-of-the-art performance in several physics-inspired metrics. We have demonstrated its capability to generate five different jet types with high fidelity using datasets consisting of 30 to 150 particles, showcasing the model's ability to generate jets with different particle multiplicities through a masking strategy.

Furthermore, the generation time was reduced by a factor of 450 using progressive distillation compared to the initial FPCD baseline, enabling high-fidelity generation with a single time step. This exciting result motivates future research to further reduce the model complexity and accelerate even further the sampling time.

The investigation of different backbone network designs is another promising direction to reduce generation time while maintaining high fidelity. The \textsc{EPiC-GAN} network structure shows great potential, with lower computational costs in higher particle multiplicity regions.

Given the flexibility of our model, we envision possible future applications in fast event generation, hadronisation models conditioned on parton kinematics, and full event reconstruction conditioned on different particle types.

\section*{Code Availability}

The code for this paper can be found at \url{https://github.com/ViniciusMikuni/GSGM}.

\section*{Acknowledgments}
We thank Jason Wong for thoughtful discussions. VM, MP, and BN are supported by the U.S. Department of Energy (DOE), Office of Science under contract DE-AC02-05CH11231. This research used resources of the National Energy Research Scientific Computing Center, a DOE Office of Science User Facility supported by the Office of Science of the U.S. Department of Energy under Contract No. DE-AC02-05CH11231 using NERSC award HEP-ERCAP0021099.  

\bibliography{HEPML,other}
\bibliographystyle{apsrev4-1}

\clearpage
\appendix

\section{Jet and particle kinematic distributions for the samples containing up to 150 particles}
\label{app:results_150}
\begin{figure*}[!htb]
    \centering
        \includegraphics[width=.35\textwidth]{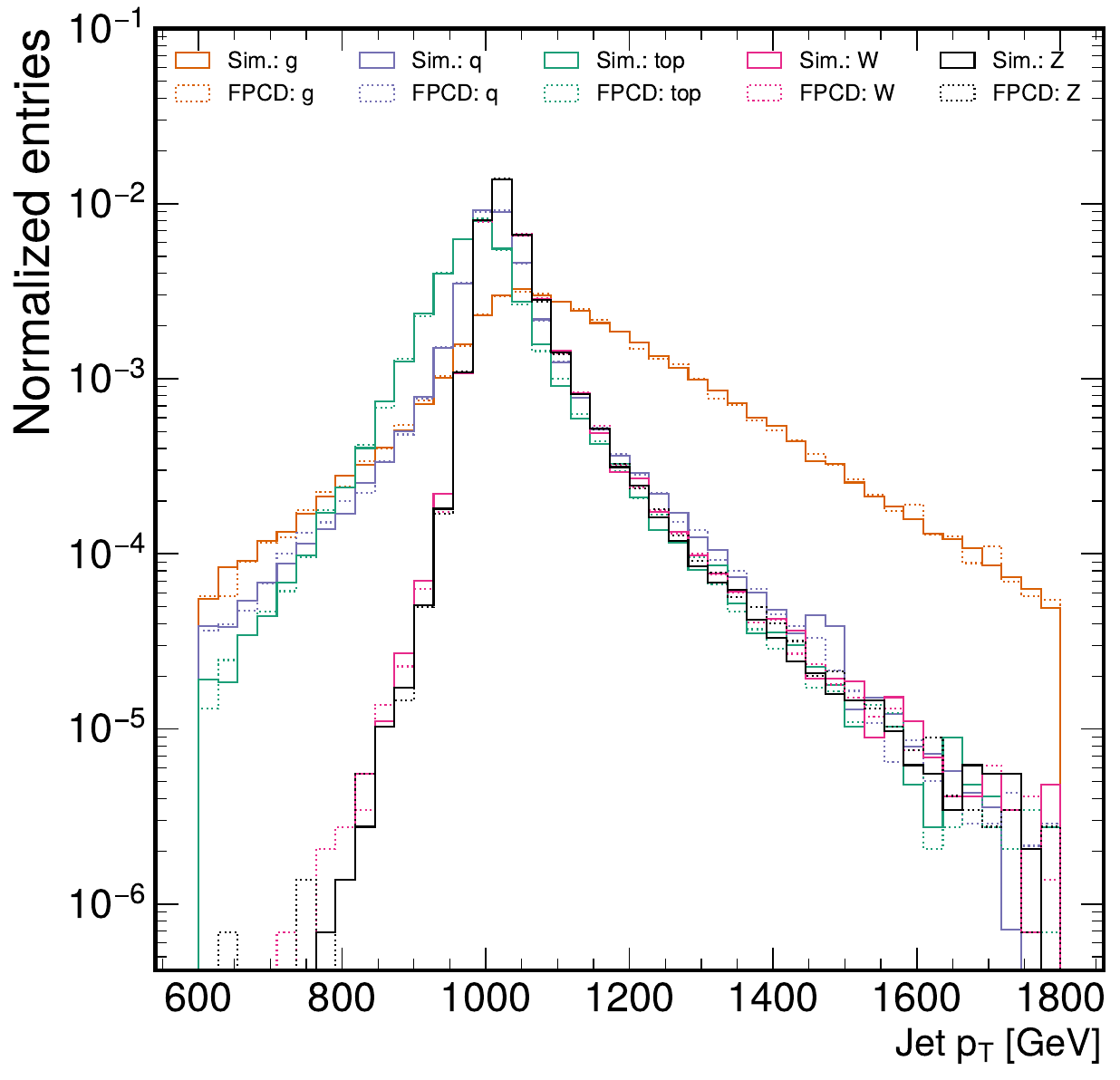}
        \includegraphics[width=.35\textwidth]{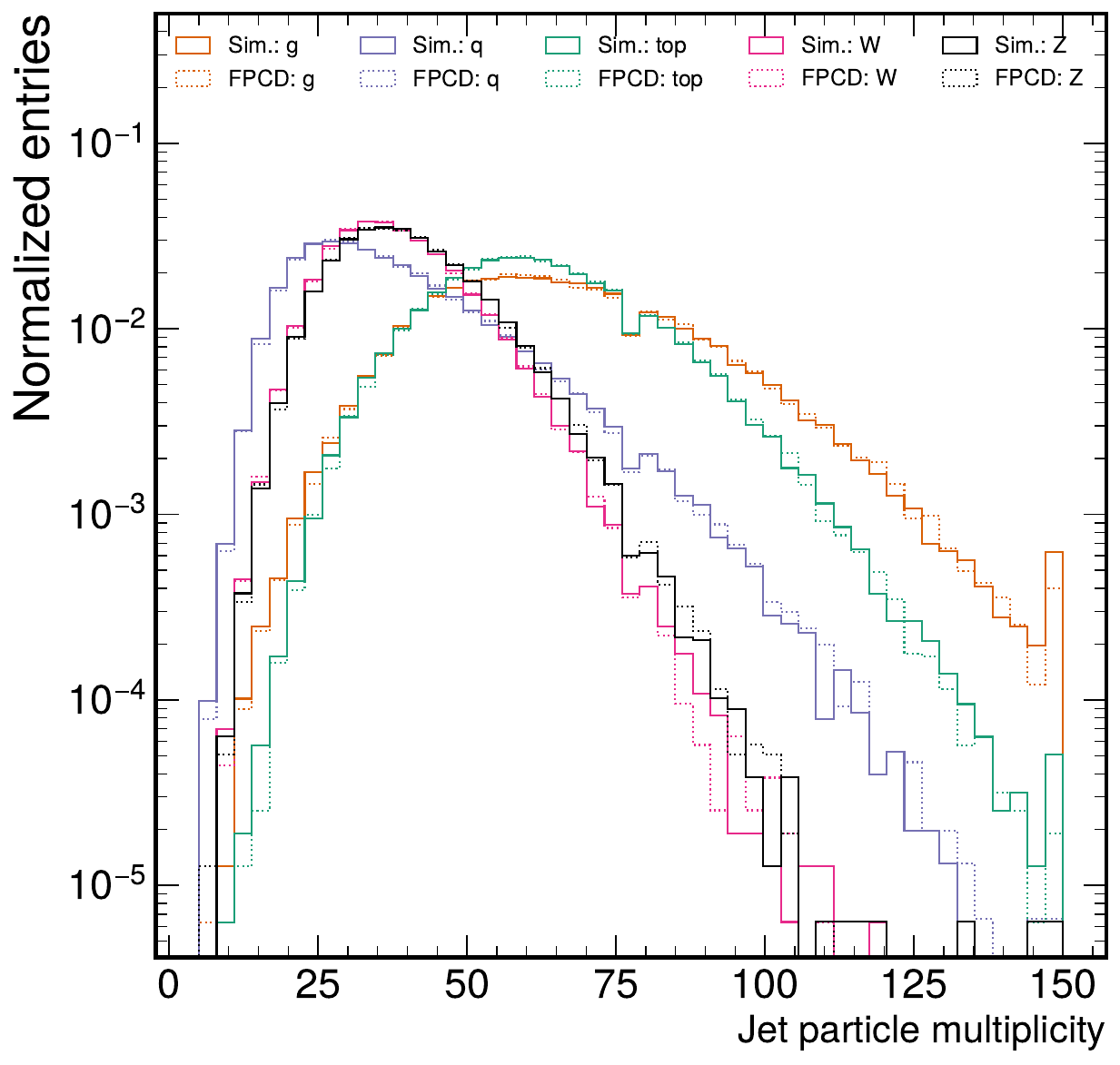} 
        \includegraphics[width=.35\textwidth]{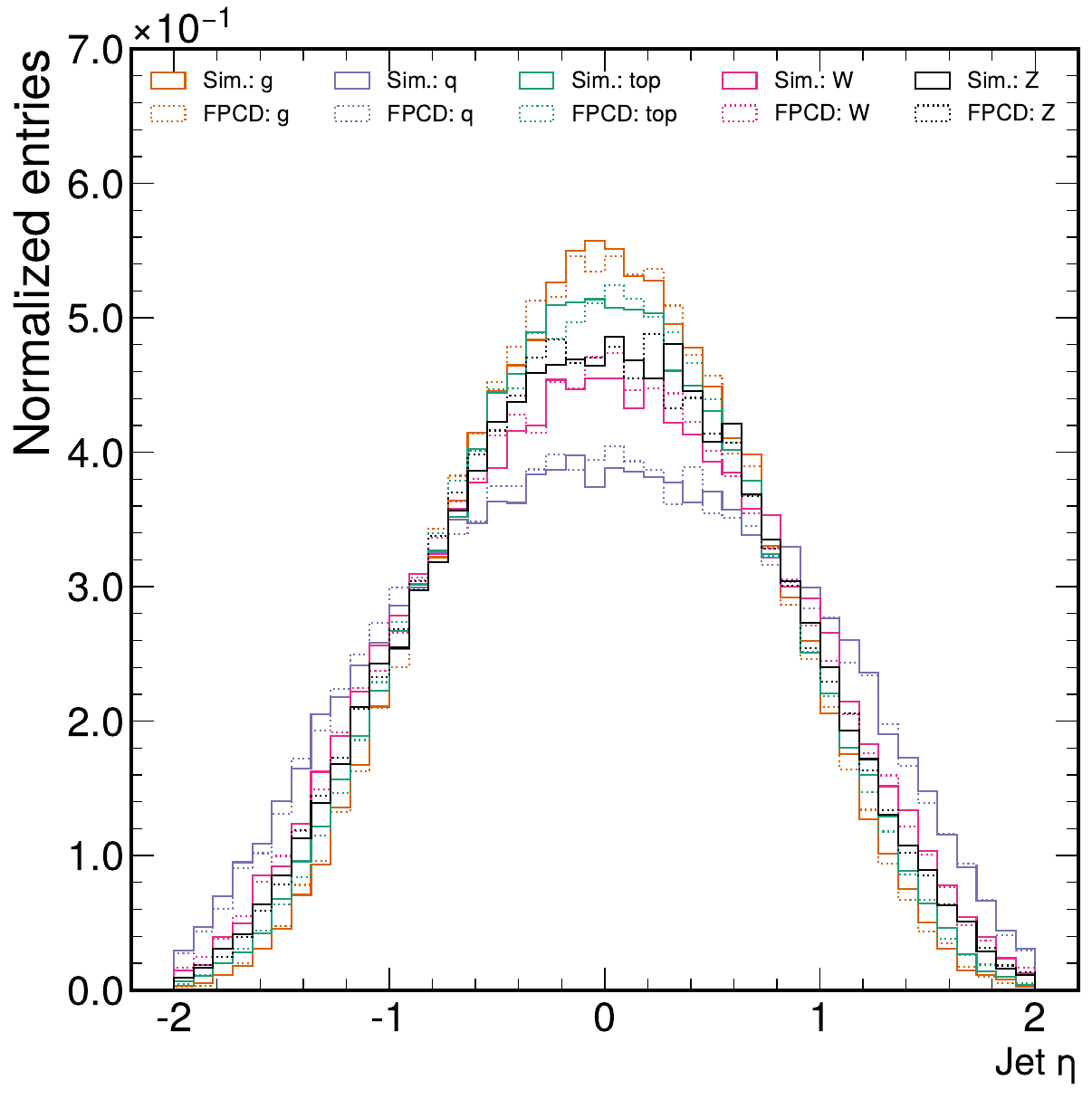}
        \includegraphics[width=.35\textwidth]{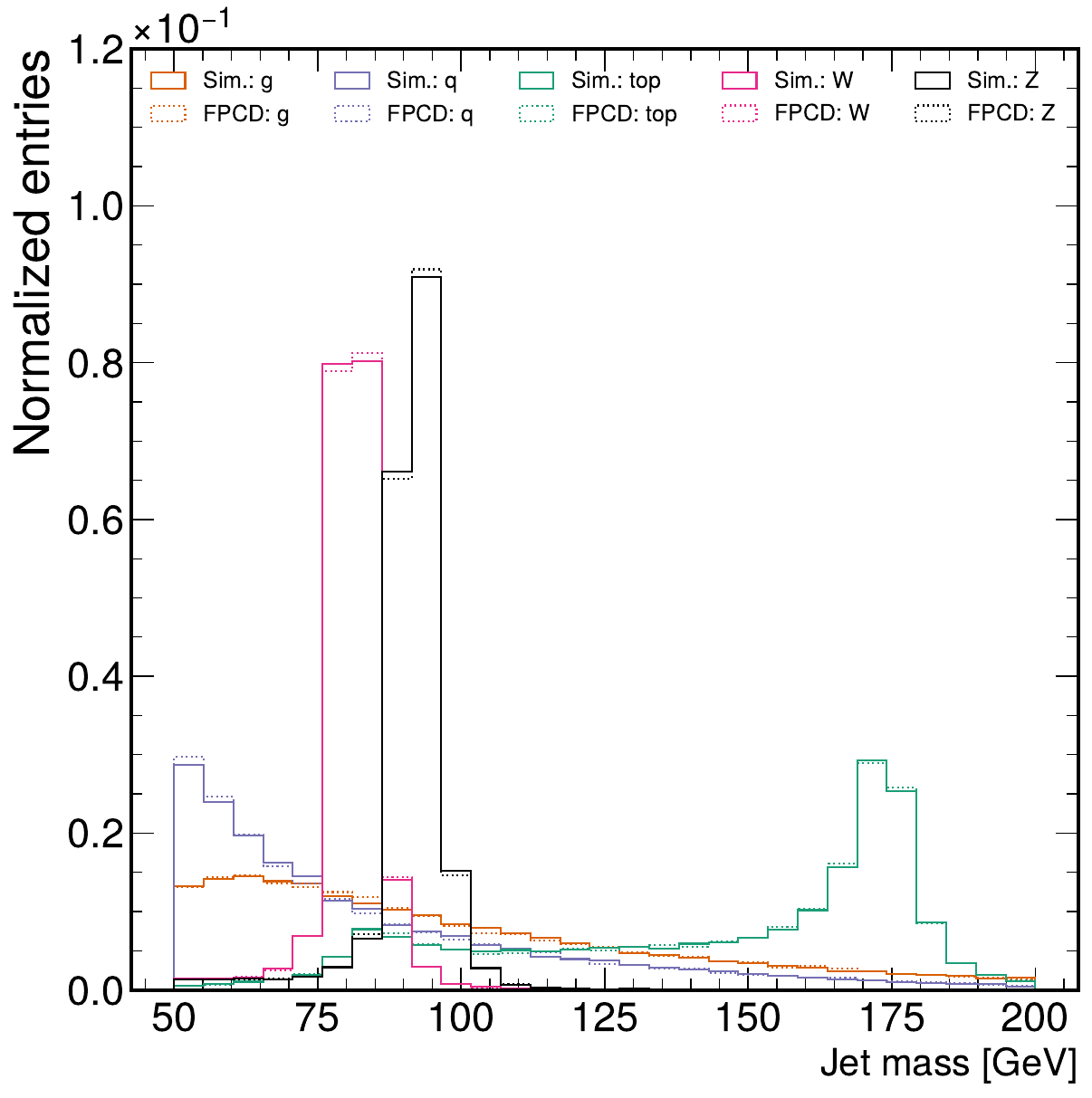}

    \caption{Generated jet kinematic information using FPCD compared to simulated events for particle jets consisting of light-quarks (q), gluons (g), and top quarks (top).}
    \label{fig:jet_150}
\end{figure*}

\begin{figure*}[!htb]
    \centering
        \includegraphics[width=.28\textwidth]{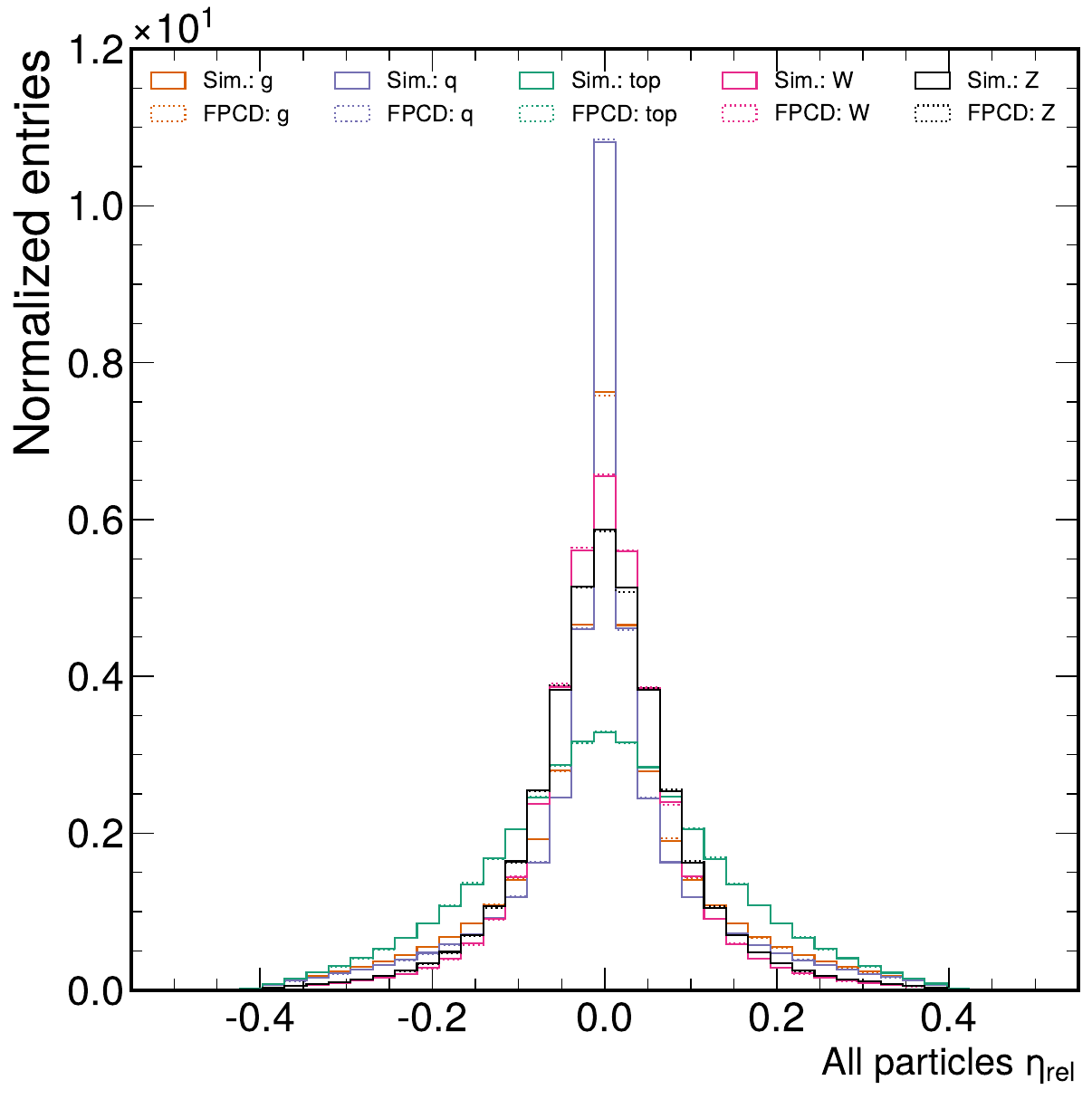}
        \includegraphics[width=.28\textwidth]{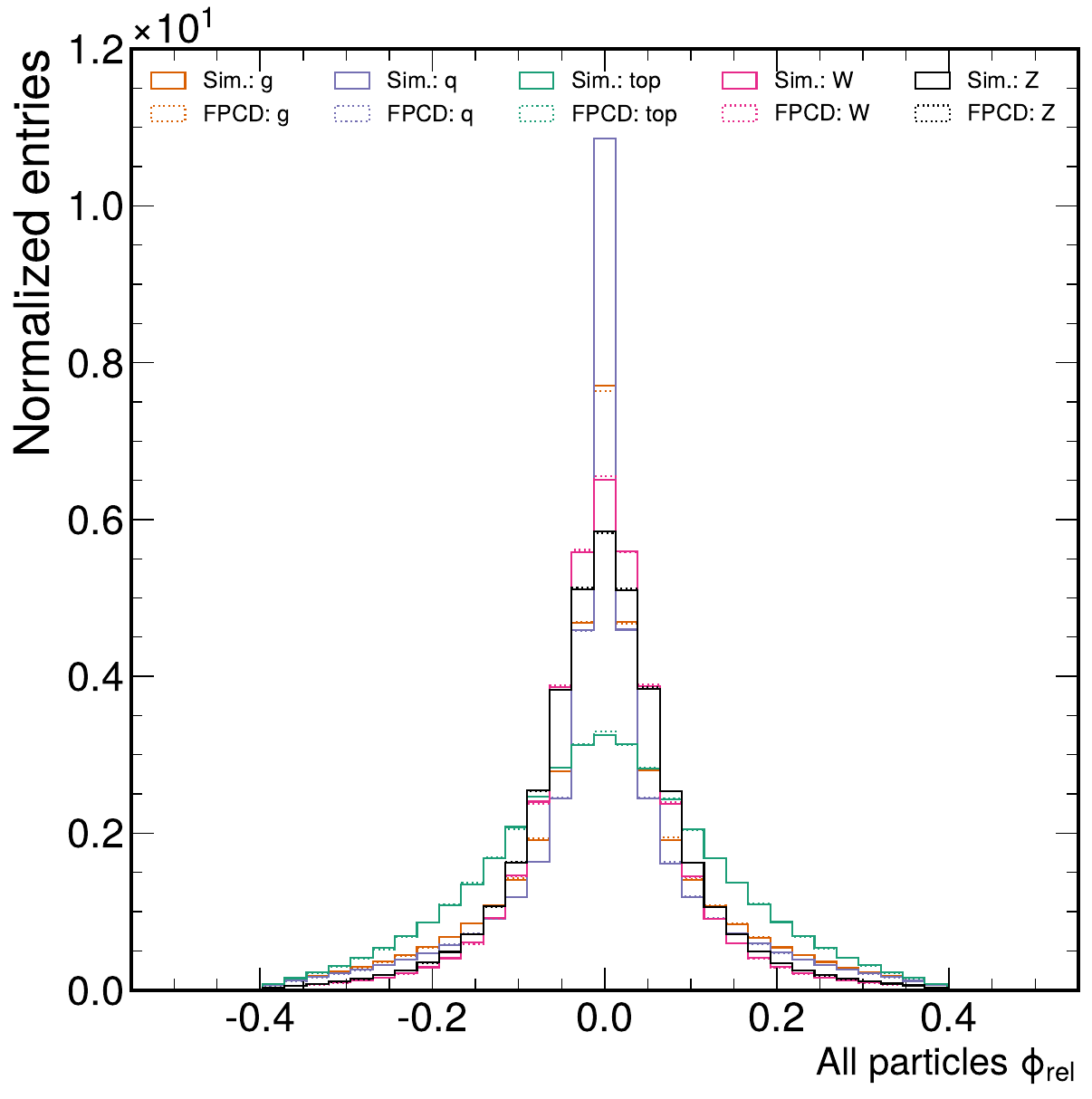}
        \includegraphics[width=.28\textwidth]{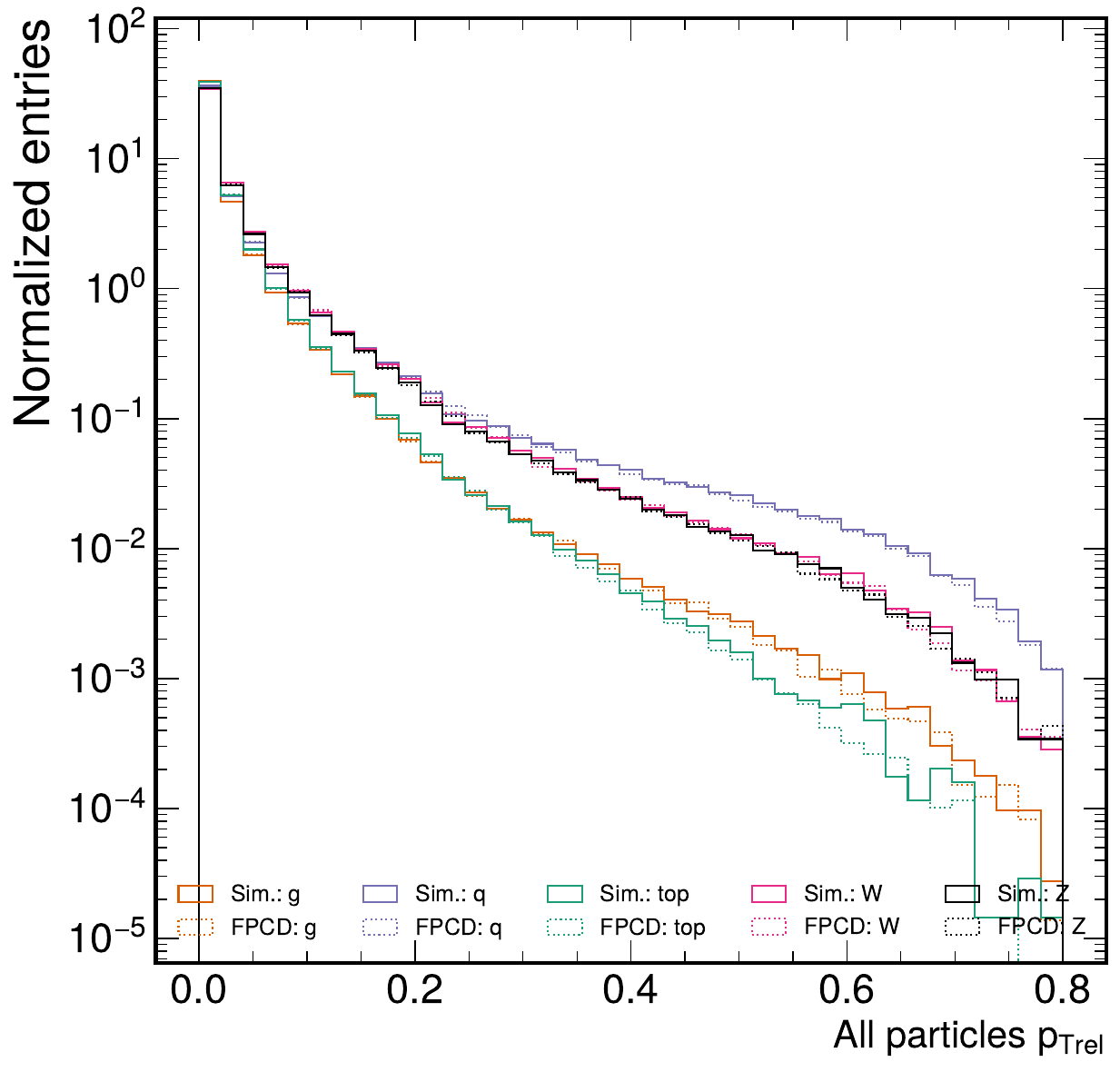}        
    \caption{Generated particle kinematic information using FPCD compared to simulated events for particle jets consisting of light-quarks (q), gluons (g), and top quarks (top). For each jet, all particles for both simulation and FPCD are shown.}
    \label{fig:part_150}
\end{figure*}

In this section we show the results obtained by the diffusion model trained using the dataset consisting of jets containing up to 150 particles in Figs.~\ref{fig:jet_150} and~\ref{fig:part_150}.

\section{Jet kinematic distributions for different distilled models}
\label{app:comp}
\begin{figure*}[!htb]
    \centering
        \includegraphics[width=.35\textwidth]{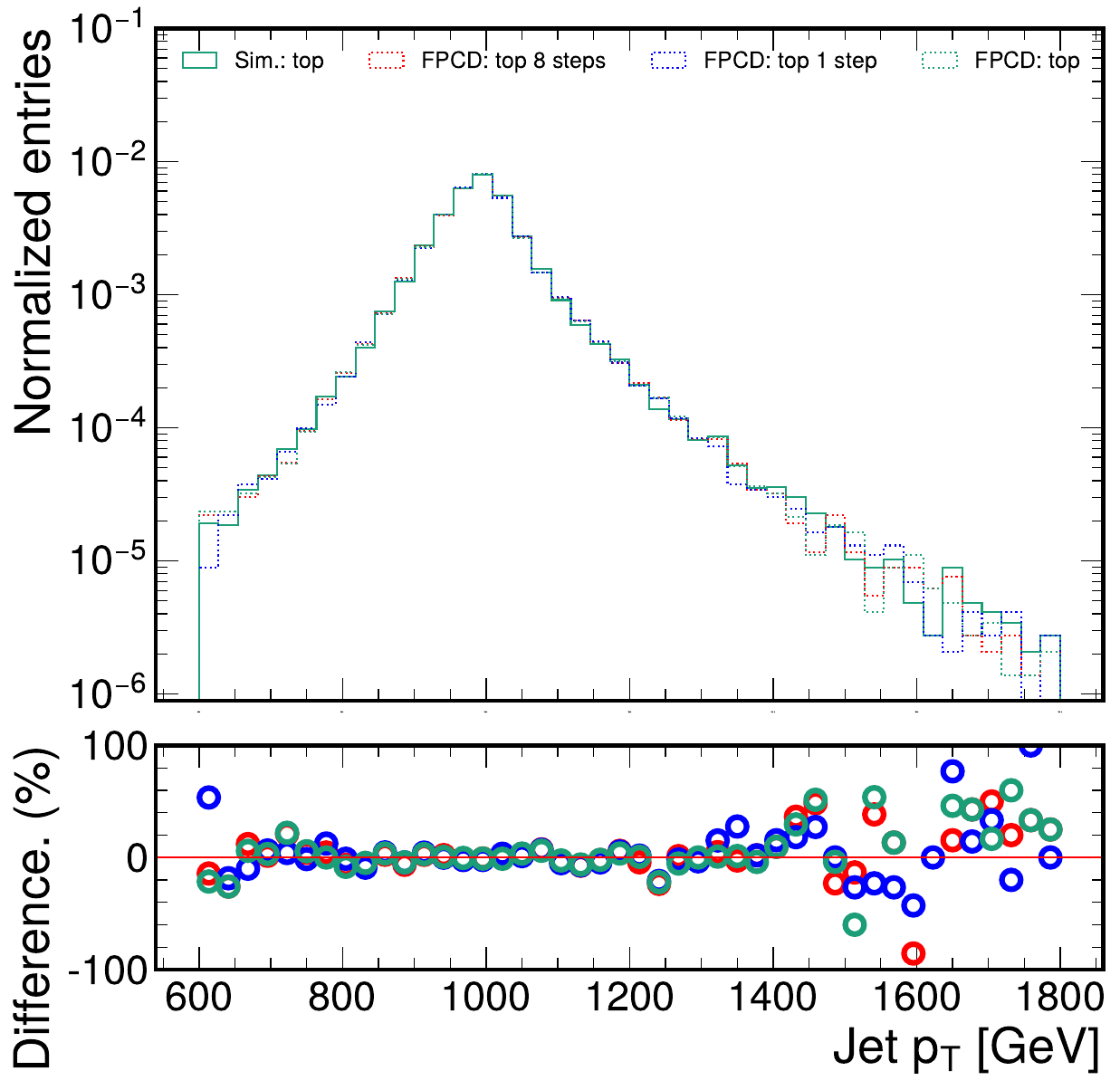}
        \includegraphics[width=.35\textwidth]{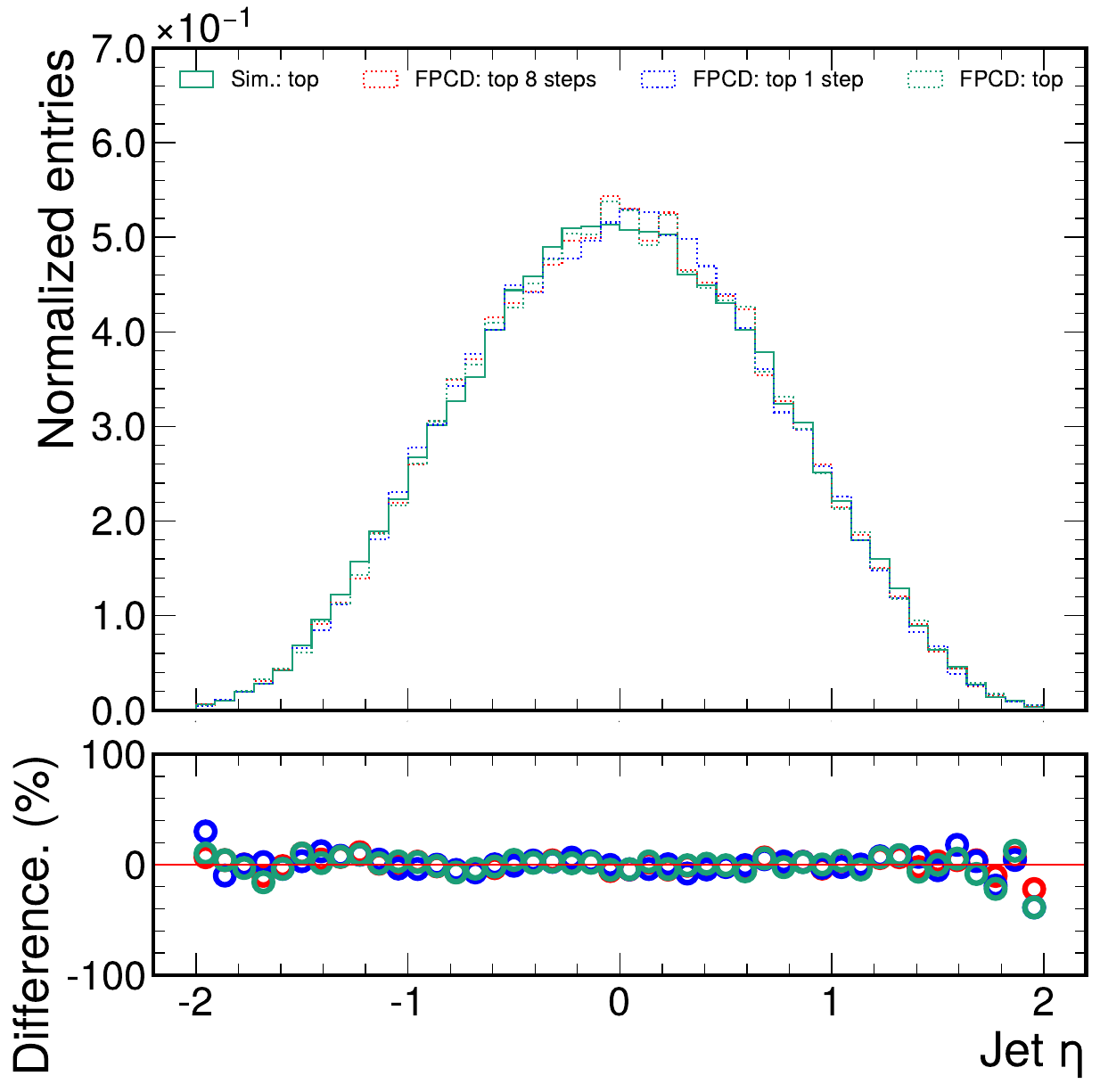}
        \includegraphics[width=.35\textwidth]{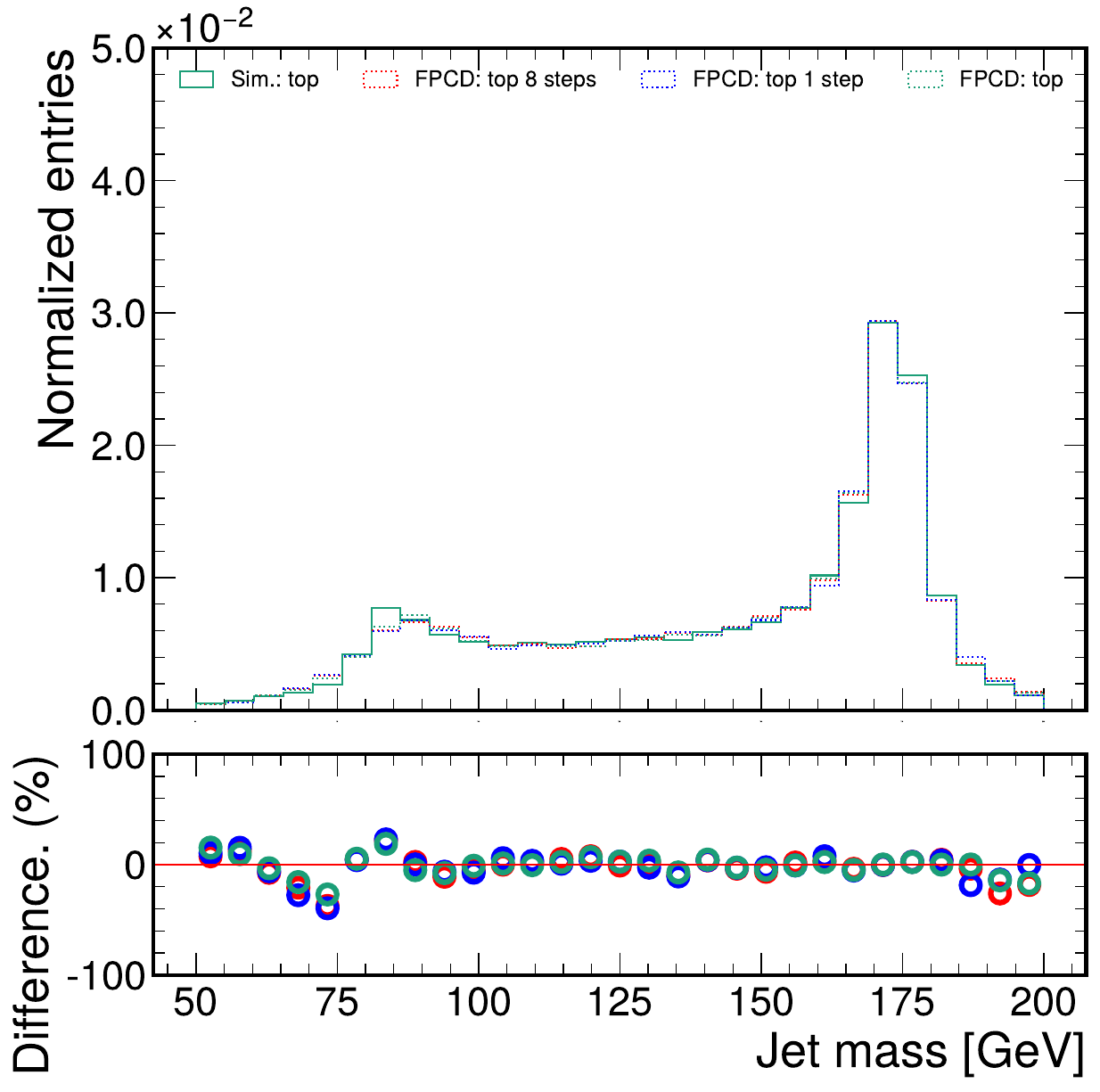}
        \includegraphics[width=.35\textwidth]{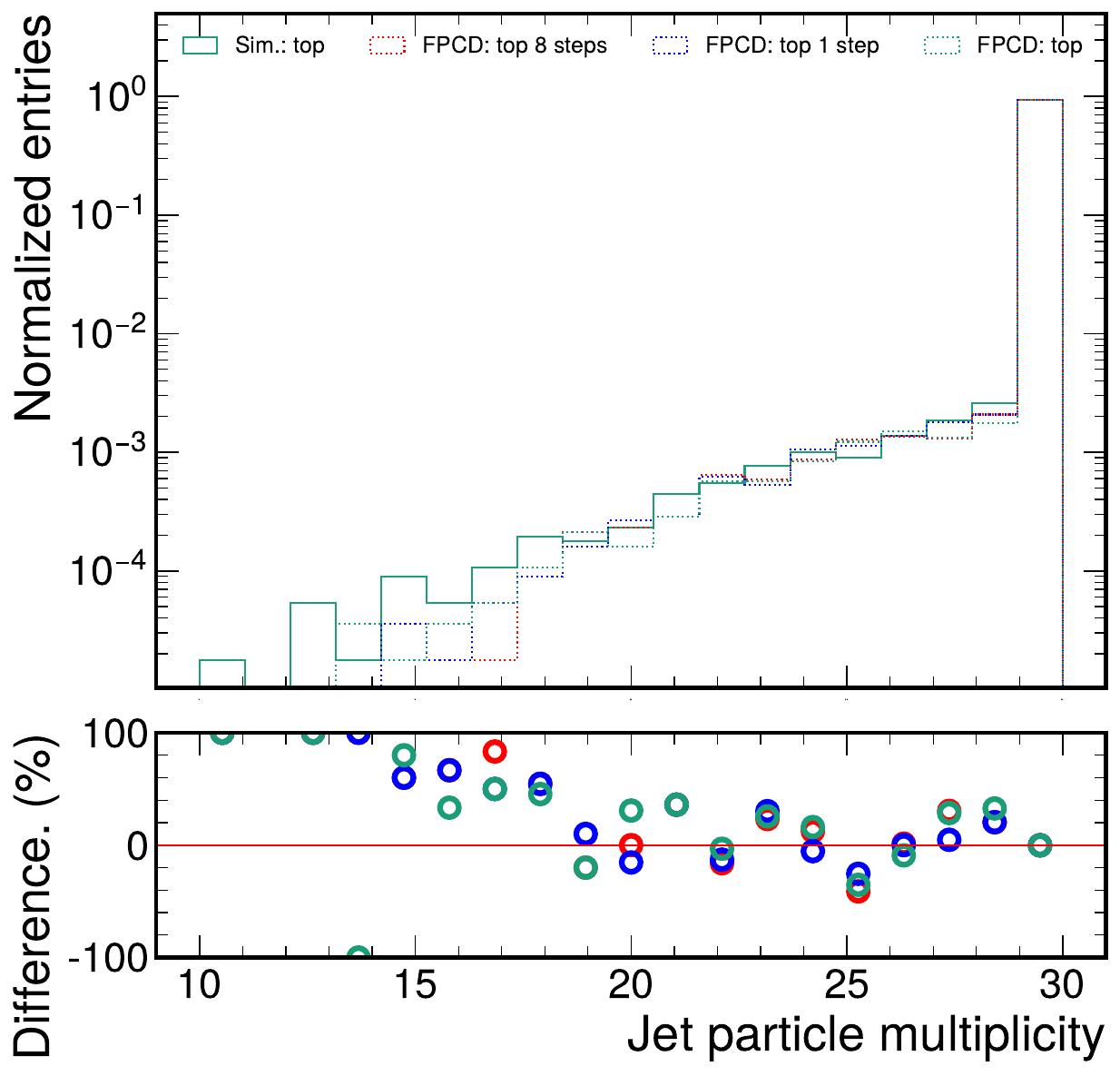}        
    \caption{Comparison of generated jet kinematic information using different distillation steps for top quark initiated jets in the dataset consisting of 30 particles.}
    \label{fig:comp_jet}
\end{figure*}

\begin{figure*}[!htb]
    \centering
        \includegraphics[width=.28\textwidth]{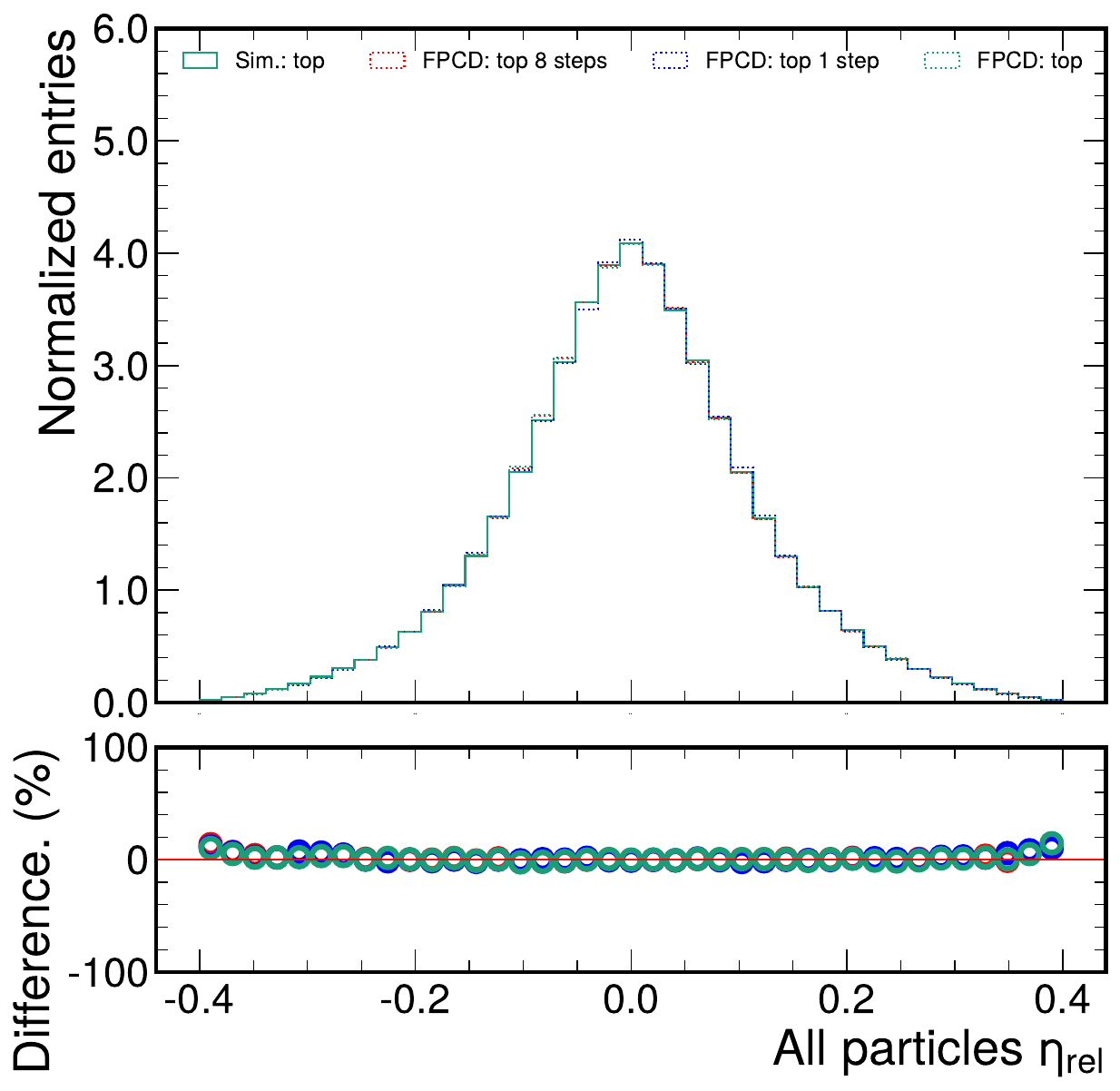}
        \includegraphics[width=.28\textwidth]{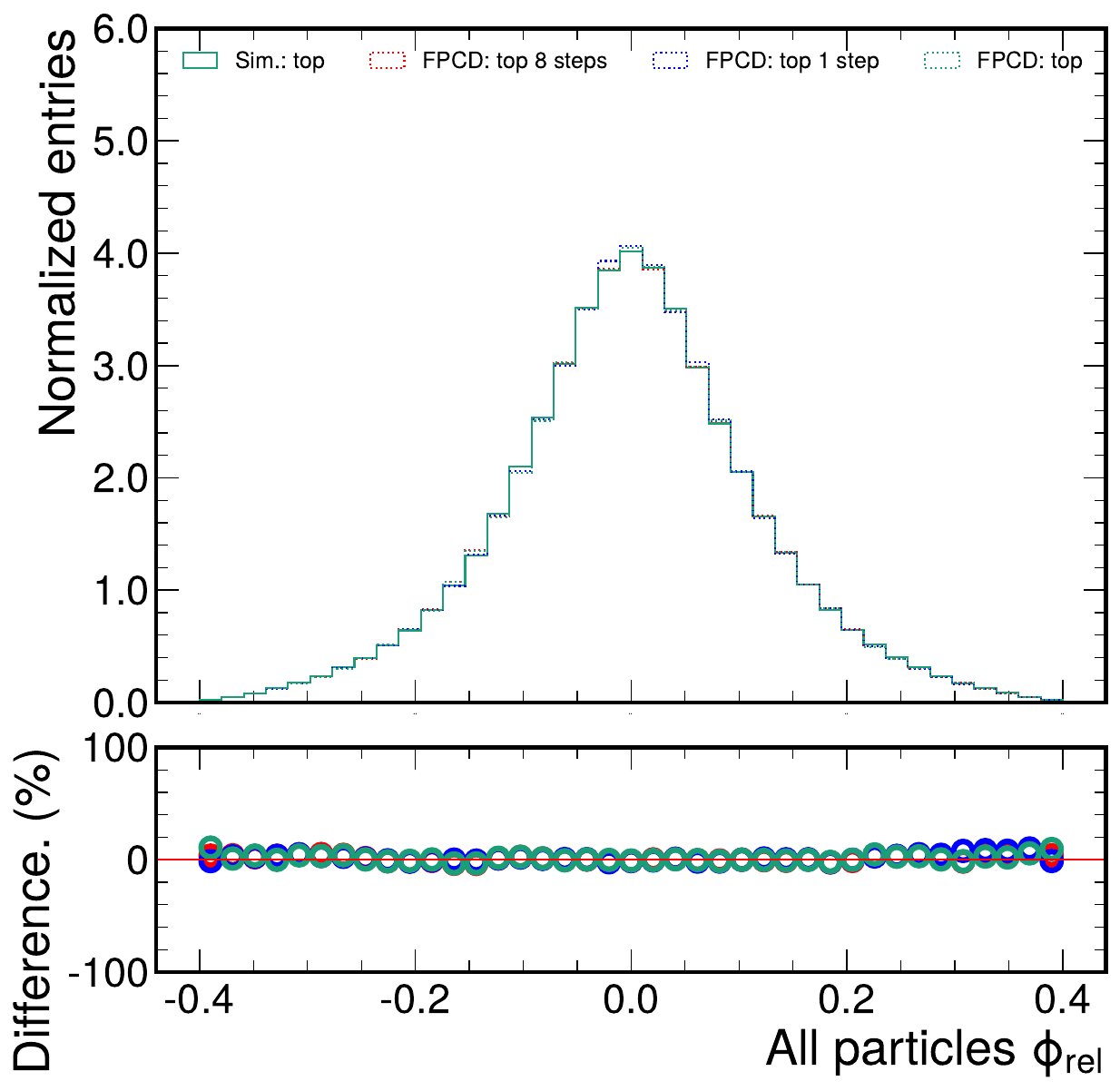}
        \includegraphics[width=.28\textwidth]{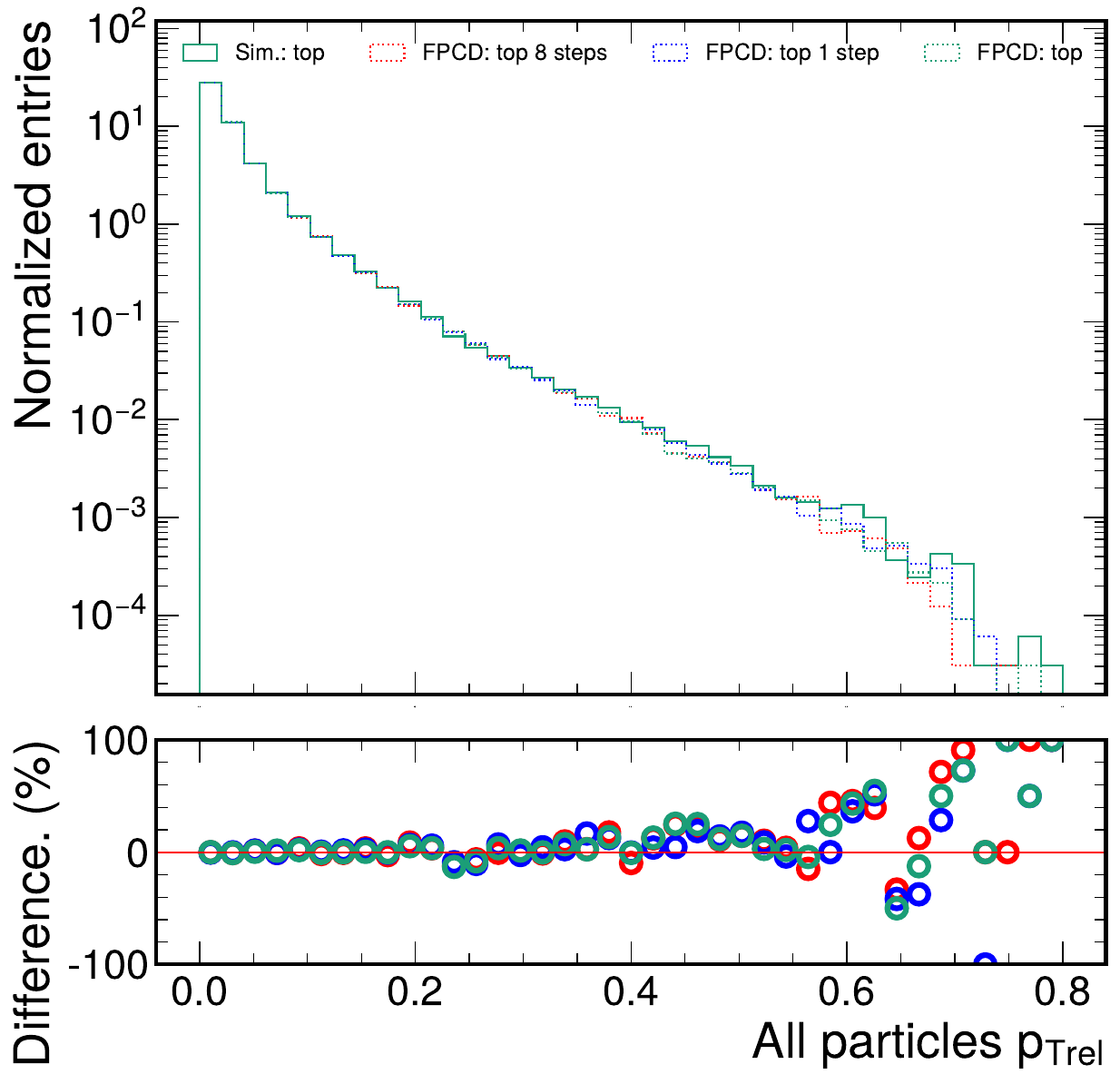}
    \caption{Comparison of generated particle kinematic information using different distillation steps for top quark initiated jets in the dataset consisting of 30 particles.}
    \label{fig:comp_part}
\end{figure*}

In this section we provide the jet and particle kinematic distributions for top quark initiated jets using different number of distillation steps. The results are shown in Fig.~\ref{fig:comp_jet} and Fig.~\ref{fig:comp_part} for jet and particle information respectively.

\end{document}